\documentclass[review]{elsarticle}
\usepackage{lineno}
\usepackage{hyperref}
\modulolinenumbers[5]
\journal{arXiv}

\usepackage{amsmath}
\usepackage{multicol}
\usepackage{graphicx}
\usepackage{adjustbox}
\usepackage{colortbl}
\usepackage{tabularx}
\usepackage{bm}
\usepackage{booktabs}
\usepackage{url}
\usepackage{subfig}
\usepackage[labelfont=bf,labelsep=period]{caption}
\usepackage[margin=1.0in]{geometry}

\begin{document}
	
	\begin{frontmatter}
		
		\title{Depth Range Reduction for 3D Range Geometry Compression}
		
		\author{Matthew G. Finley}
		\author{Tyler Bell\corref{mycorrespondingauthor}}
		
		\cortext[mycorrespondingauthor]{Corresponding author}
		\ead{tyler-bell@uiowa.edu}
		
		\address{Department of Electrical and Computer Engineering, The University of Iowa, Iowa City, IA 52242, USA}
		
		\begin{abstract}
			
			Three-dimensional (3D) shape measurement devices and techniques are being rapidly adopted within a variety of industries and applications.
			As acquiring 3D range data becomes faster and more accurate it becomes more challenging to efficiently store, transmit, or stream this data.
			One prevailing approach to compressing 3D range data is to encode it within the color channels of regular 2D images.
			This paper presents a novel method for reducing the depth range of a 3D geometry such that it can be stored within a 2D image using lower encoding frequencies (or a fewer number of encoding periods).
			This allows for smaller compressed file sizes to be achieved without a proportional increase in reconstruction errors.
			Further, as the proposed method occurs prior to encoding, it is readily compatible with a variety of existing image-based 3D range geometry compression methods.
			
		\end{abstract}
		
		\begin{keyword}
			Depth range reduction\sep 3D range data compression\sep Depth encoding\sep Fringe analysis
		\end{keyword}
		
	\end{frontmatter}

\section{Introduction}
\label{sec:introduction}

Three-dimensional (3D) shape measurement devices and techniques have improved greatly in recent decades.
There is now a robust ecosystem of options available with which to acquire 3D range data, including stereo vision, time-of-flight, and structured light.
In general, these methods are capturing 3D range data at ever increasing speeds and with greater accuracies.
Alongside these advantages, a variety of industries and applications are adopting 3D imaging devices and techniques, including the forensic sciences, manufacturing, entertainment, and medicine.
Progress, however, does not come without challenges.
Faster and more accurate 3D range data acquisition equates to larger quantities of data that must be stored with greater fidelity.
This challenge becomes more complex when attempting to efficiently transmit or stream the high-fidelity 3D range data, as is desired for applications like telemedicine and telepresence.

A traditional method of representing 3D data is with a mesh data structure. A mesh stores a collection of vertices that specify 3D coordinate positions and edges; these edges specify how the vertices are connected to each other.
There are several popular 3D mesh file formats, including OBJ, STL, and PLY.
Depending on which one is utilized, it is common to store additional data with a mesh, such as surface normals, vertex colors, texture coordinates, and texture maps. 
To make 3D mesh storage more efficient, much work has been done to compress 3D meshes~\cite{Peng:MeshCompressionSurvey:2005,Maglo:3DMeshCompressionSurvey:2015}. 
Generally, mesh compression methods require two steps: connectivity compression and geometry compression.
Connectivity compression methods~\cite{Deering:GeometryCompression:1995,Taubin:TopologicalSurgery:1998,Touma:TriangleMeshCompression:1998,Gumhold:TriangleMeshConnectivity:1998,Rossignac:Edgebreaker:1999,Alliez:ValenceDrivenConnectivityEncoding:2001} attempt to define a simplified traversal of the mesh's vertices such that the edges can be represented with less information than the original.
Geometry compression methods attempt to reduce the amount of information used to represent the mesh's 3D coordinates, typically achieved through a process of quantization, prediction, and entropy encoding~\cite{Peng:MeshCompressionSurvey:2005,Maglo:3DMeshCompressionSurvey:2015}.

When an application can afford to alter a mesh's original structure, greater efficiency can be achieved through mesh remeshing (i.e., resampling) and simplification (i.e., downsampling)~\cite{Botsch:PolygonMeshProcessing:2010}. 
The goal of such techniques is to generate a mesh that is, to varying degrees of loss, faithful to the original mesh.
The ideal result is a mesh with a reduced footprint---due to less required connectivity and geometry information---that functionally looks or performs the same as the original mesh.
File sizes can be further reduced if the mesh itself has an underlying structure that inherently represents the connectivity information, such as the regular grid or pixel-like structure present in 3D range data derived from 2D imaging devices.
In these cases, the embedded connectivity structure can be used to produce a smaller representation of the geometry data as the connectivity information does not have to be explicitly stored or compressed. 

An approach that takes advantage of the aforementioned inherent grid-like structure is to encode 3D coordinates within the color pixels of a regular 2D image.
One such technique to store 3D data in a 2D image is to use a virtual structured light scanner~\cite{Karpinsky:Holoimage:2010,Karpinsky:Holovideo:2012,Hou:TwoChannel:2012,Karpinsky:H264:2013,Karpinsky:3Bit:2013,Wang:TwoChannel:2016}.
These image-based methods utilize principles of phase-shifting with a virtual camera and virtual projector to create digital 2D encodings of the 3D surface being compressed.
Once the encodings have been produced they can then be stored in the color channels of a 2D image and traditional image compression techniques (e.g., PNG, JPEG) can be leveraged to achieve even smaller file sizes.

Another approach to compressing 3D range data is to create 2D encodings based directly on the depth map, $Z$. 
Methods that follow this approach~\cite{Zhang:DirectDepth:2012,Ou:SMap:2013,Bell:MWD:2015,Finley:TCD:2019} directly encode the 3D range data's floating-point depth map using principles of phase-shifting, such that one or more high-frequency 8-bit encodings are produced.
These encodings are then stored in one or two of the output 2D image's color channels.
One additional channel of the 2D image is then typically used to store auxiliary information necessary for the eventual recovery of the 3D geometry information stored within the high-frequency encoded channels. 
As before, once the encodings have been generated and placed into the color channels of a regular 2D image, the image can be compressed with traditional image compression techniques.

In general, image-based 3D range geometry compression methods are computationally efficient, can recover 3D geometry with low reconstruction error, and result in high compression ratios.
Even so, there have been attempts to improve the file savings of such methods further. 
For example, Zhang et al.~\cite{Zhang:HoloimageInterpolation:2016} proposed a method that performed image downsampling on the 2D encoded output image to further reduce its file size.
It was the downsampled, lower-resolution 2D image that represented the original geometry and was stored.
To recover the 3D range geometry, their method upsampled the 2D image to its original dimensions and then decoded the data stored within its color channels. 
The work of Zhang et al. work illustrates that pre- and post-processing techniques can also contribute to an overall increase in efficiency for image-based 3D geometry compression.

Although the previously mentioned depth compression methods each vary in their specific approach to encoding the depth map, $Z$, it was realized that they are each heavily influenced by the overall depth range of the geometry to be encoded.
This paper presents a novel method of reducing the depth range of a 3D range geometry such that it can be encoded using lower encoding frequencies (i.e., with fewer encoding periods).
As the reduction in depth range allows for a proportional reduction in encoding frequency, this allows for smaller compressed file sizes to be achieved without a subsequent increase in reconstruction error.
Further, as the proposed depth range reduction method occurs prior to encoding, it is readily compatible with a variety of existing image-based 3D range geometry compression methods.

The remainder of this paper is organized as follows: Sec.~\ref{sec:principle} details the theory and working principles of the proposed depth range reduction method; Sec.~\ref{sec:experiments}  validates the proposed method via a variety of experimental results; Sec.~\ref{sec:discussion} discusses some of the practical challenges that face the proposed method; Sec.~\ref{sec:conclusion} summarizes and concludes the paper.

\section{Principle}
\label{sec:principle}
Introduced in Sec.~\ref{sec:introduction}, one technique to compress 3D range geometry is to encode it within the RGB color channels of a 2D image.
Specifically, approaches that encode the depth data (or a similar equivalent~\cite{Ou:SMap:2013}) directly have been shown to offer large compression ratios while being robust to various 2D image compression formats.
Typically, 3D depth compression methods employ principles of phase-shifting to produce one or more sinusoidal 2D encoding(s) of the 3D range data.
An additional encoding  that contains some auxiliary information necessary to recover the data stored within the sinusoidal encoding(s) is also usually produced.
After the encoding process, the sinusoidally encoded image(s) and the auxiliary encoding can be placed within the color channels of a 2D image that can be compressed with methods such as PNG or JPEG.

One example of these depth encoding methods is the multiwavelength depth (MWD) encoding method~\cite{Bell:MWD:2015}.
This method encodes a floating-point depth map, $Z$, into the three 8-bit color channels of a 2D output image.
Mathematically, this method's encodings are generated as follows:
\begin{equation}
I_{r}(i,j) = \frac{1}{2} + \frac{1}{2} \sin \left( 2\pi \times \frac{Z(i,j)}{P} \right),
\label{eqn:mwd:r}
\end{equation}
\begin{equation}
I_{g}(i,j) = \frac{1}{2} + \frac{1}{2} \cos \left( 2\pi \times \frac{Z(i,j)}{P} \right),
\label{eqn:mwd:g}
\end{equation}
\begin{equation}
I_{b}(i,j) = \frac{Z(i,j) - \min( Z )}{\text{Range}(Z)}.
\label{eqn:mwd:b}
\end{equation}
In these equations, $(i,j)$ denotes pixel indicies, $min$ is the function that finds the minimum pixel value in an image, and $Range$ is the function that returns the absolute difference between between the maximum and minimum values in an image.
The value $P$ is a user-defined \emph{fringe width} which specifies the depth distance in the range of $Z$ to be encoded within each period of Eq.~(\ref{eqn:mwd:r}) and~(\ref{eqn:mwd:g}).
To encode $Z$ into $n$ equal periods, $P$ is defined as:
\begin{equation}
P = \frac{\text{Range}(Z)}{n}.
\label{eqn:mwd:p}
\end{equation}
In general, as the number of periods ($n$) used to encode values within a depth range increase, $Z$ is encoded at a higher frequency; this leads to more precise reconstructions but larger file sizes due to the increased rate of intensity variation within the 2D output image.
Conversely, as the number of periods used to encode values within a depth range decrease, $Z$ is encoded at a lower frequency; this leads to less precise reconstructions but smaller compressed file sizes.  

\subsection{Encoding with Depth Range Reduction}
\label{sub:principle-encoding}

This paper proposes a method for reducing the range of values present in the geometry to be encoded, allowing for fewer encoding periods to be used to encode the geometry.
This leads to smaller compressed file sizes without the typical proportional increase in reconstruction errors associated with a reduced number of encoding periods.

To achieve a reduced depth range in the 3D geometry to be encoded, an approximated geometry can be generated and subtracted from the original geometry. This subtraction can be defined mathematically as:
\begin{equation}
Z_{r}(i,j) = Z(i,j) - \widetilde{Z}'(i,j).
\label{eqn:drr:subtraction}
\end{equation}
Here, $Z$ is the original 3D range geometry's depth map; $\widetilde{Z}'$ is a pixel-aligned approximation of the original depth map, $Z$; and $Z_r$ is a new range-reduced depth map to be encoded using any of the referenced depth-based compression methods.
For example, $Z_r$ could be immediately compressed with the MWD encoding approach by substituting $Z_r$ for every instance of $Z$ in Eq.~(\ref{eqn:mwd:r})-(\ref{eqn:mwd:b}).
As $Z_r$ ideally has a smaller depth range (i.e., $Range(Z_r) < Range(Z)$), the same fringe width value of $P$ in Eq.~(\ref{eqn:mwd:p}) can be achieved with a proportionally lower number of encoding periods, $n$.
This reduction in depth range, and the subsequent reduction in the number of encoding periods required, results in a relatively lower encoding frequency.
This lower-frequency encoded output requires fewer bytes to store, while also maintaining the same reconstruction accuracy achieved by a higher-frequency encoding of the original geometry.

Although the expression used in Eq.~(\ref{eqn:drr:subtraction}) to generate the range-reduced depth map only involves a simple subtraction, the generation of the pixel-aligned approximate depth map, $\widetilde{Z}'$, is a crucial aspect of this method.

\subsubsection{Approximation}
\label{sub:principle-approximation}

Let $M$ define the original 3D range data input to the proposed method, with vertices corresponding to points within the geometry's $X$, $Y$, and $Z$ maps.
To reduce the depth range, an approximated 3D geometry, $\widetilde{M}$, must be generated such that its corresponding depth map, $\widetilde{Z}$, can be subtracted from $M$'s depth map, $Z$, to produce the range-reduced depth map, $Z_r$.
In general, the more closely the approximation's depth map ($\widetilde{Z}$) resembles the original geometry ($Z$), the more the resulting depth range will be reduced.
However, since the approximated depth map will be required by the decoding process in order to recover the original depth map, the parameters necessary to generate $\widetilde{Z}$ must be stored alongside the depth-encoded representation of $Z_r$.
It is therefore important to generate an approximation ($\widetilde{Z}$) that requires as little information as possible while maintaining the ability to reduce the depth range of the original geometry ($Z$).
The additional storage cost introduces a new variable that can be adjusted to tune this method to the target application: increasing the number of parameters required to generate $\widetilde{Z}$ can result in a higher-quality approximation but will reduce the file size savings that can be achieved through the use of the proposed method.
To generate $\widetilde{Z}$, 3D operations~\cite{Botsch:PolygonMeshProcessing:2010} or 2D operations~\cite{Sonka:ImageProcessing:2015} can be performed on $M$ or $Z$, respectively.
The specific methods that were used to generate $\widetilde{Z}$ are shown and discussed in Fig.~\ref{fig:principle-drr-breakout} and Sec.~\ref{sec:experiments}.

\subsubsection{Alignment}
\label{sub:principle-alignment}
Once the 3D range geometry approximation $\widetilde{M}$ has been generated it may be necessary to align it to the original 3D geometry, $M$.
In theory, the better aligned the two geometries are the more the resulting depth range of $Z_r$ can be reduced.
To align the two geometries, a set of common 3D transformations can be applied to each 3D coordinate, $\begin{bmatrix}\tilde{x} & \tilde{y} & \tilde{z} & 1\end{bmatrix}^t$, of the approximated 3D geometry, $\widetilde{M}$, to produce the aligned approximated 3D geometry, $\widetilde{M}'$.
Mathematically, the alignment can be produced through a linear transformation:
\begin{equation}
\begin{bmatrix}\tilde{x}' & \tilde{y}' & \tilde{z}' & 1\end{bmatrix}^t = \mathbf{T} \begin{bmatrix}\tilde{x} & \tilde{y} & \tilde{z} & 1
\end{bmatrix}^t.
\label{eqn:drr:transformation}
\end{equation}
In Eq.~(\ref{eqn:drr:transformation}), $\mathbf{T}$ is a $4 \times 4$ transformation matrix that is computed through a combination of commonly used transformations.
Here, the transformation matrix can be computed as
\begin{equation}
	\mathbf{T}=\mathbf{Translation} \times \mathbf{Rotation} \times \mathbf{Scale},
\end{equation}  
where $\mathbf{Translation}$, $\mathbf{Rotation}$, and $\mathbf{Scale}$ are $4 \times 4$ matrices that specify the parameters of their corresponding operations. 

The proposed method is independent of the approach used to estimate the transformation used in alignment; it can even be arbitrarily defined.
To algorithmically determine a more precise transformation, mesh registration approaches~\cite{Torr:MLESAC:2000, Tam:3DRegistration:2012} can be employed. 
In the simplest case, the approximated geometry is derived from the 3D geometry ($M$) or from the depth map ($Z$) directly; thus, the approximation is already well aligned to the original geometry.
In this case, $\mathbf{T}$ is simply a $4 \times 4$ identity matrix.

\begin{figure}[h!]
	\centering
	\subfloat[\label{fig:principle-drr-breakout-depth}]{\includegraphics[width=0.24\columnwidth]{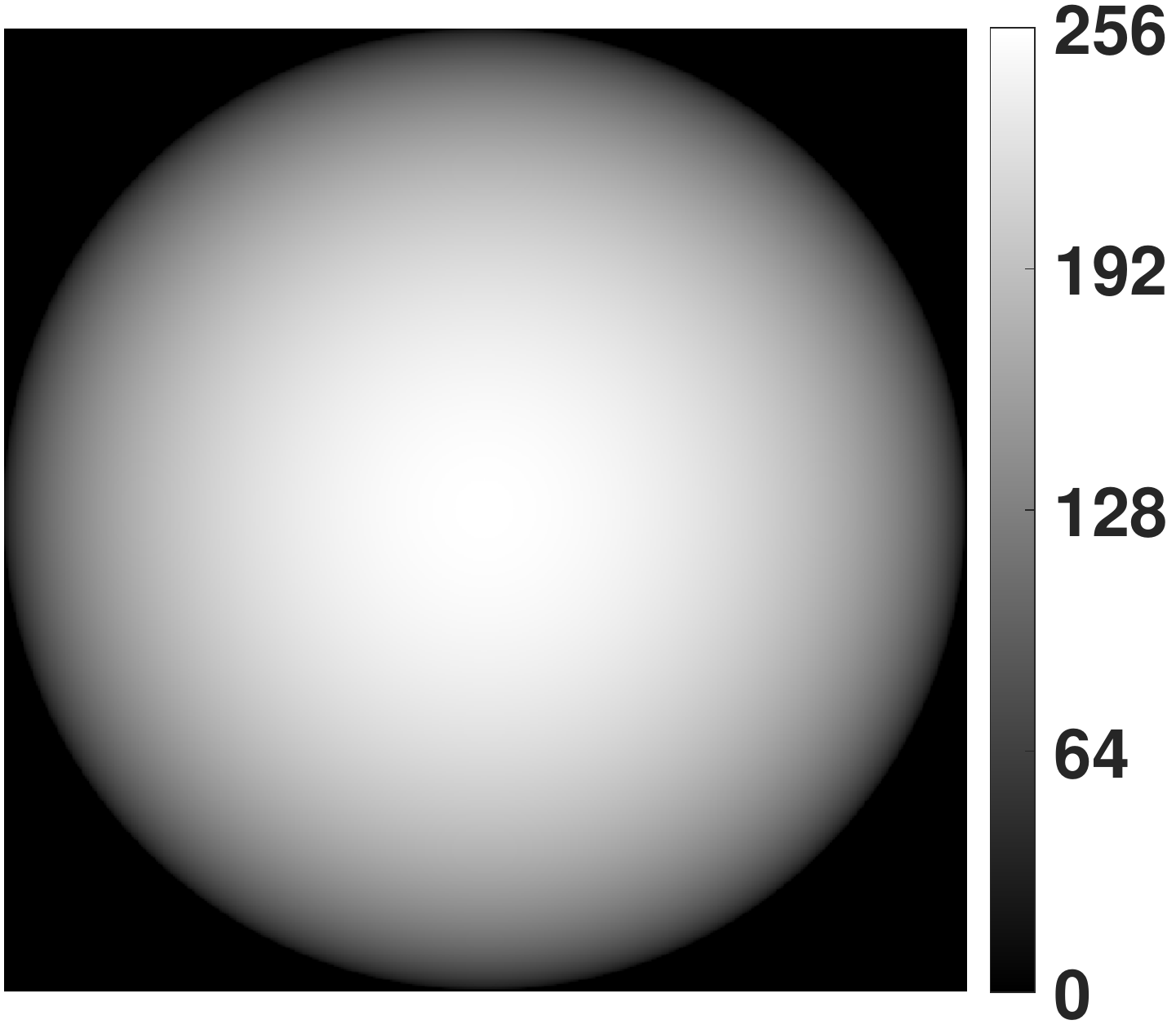}}
	\hspace{0.3em}
	\subfloat[\label{fig:principle-drr-breakout-depth-approx}]{\includegraphics[width=0.24\columnwidth]{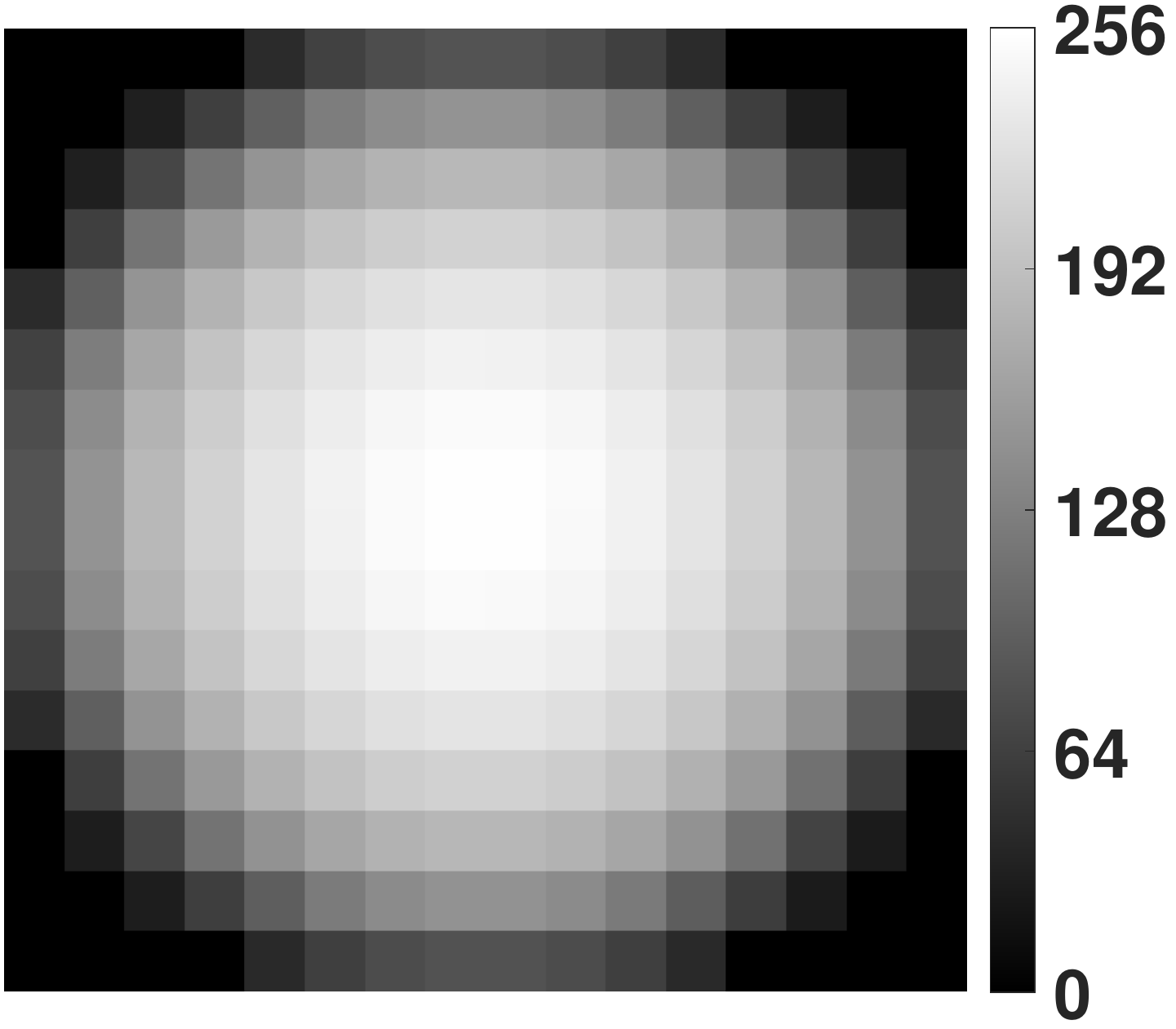}}
	\hspace{0.3em}
	\subfloat[\label{fig:principle-drr-breakout-depth-approx-filtered}]{\includegraphics[width=0.24\columnwidth]{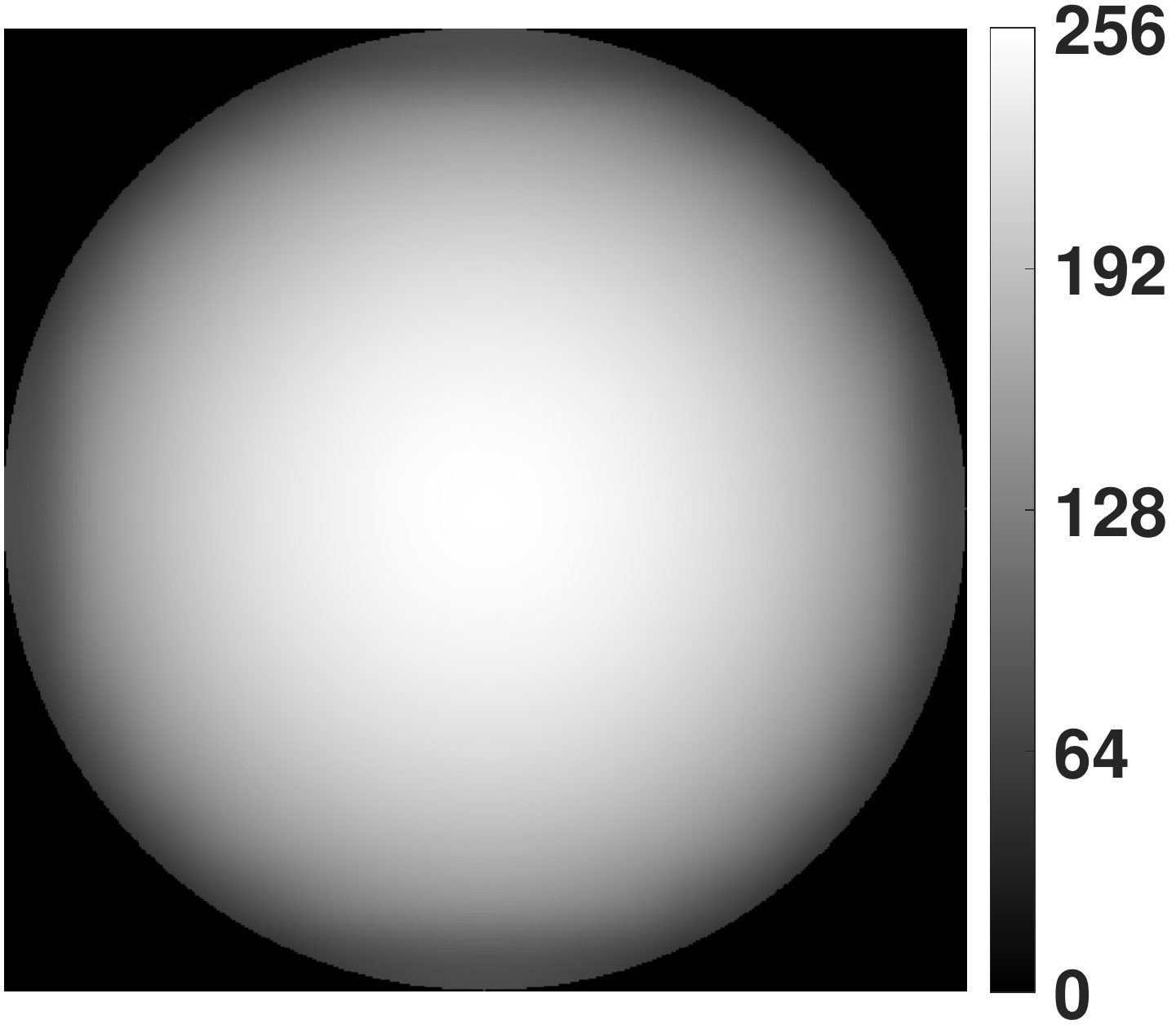}}
	\hspace{0.3em}
	\subfloat[\label{fig:principle-drr-breakout-depth-reduced}]{\includegraphics[width=0.24\columnwidth]{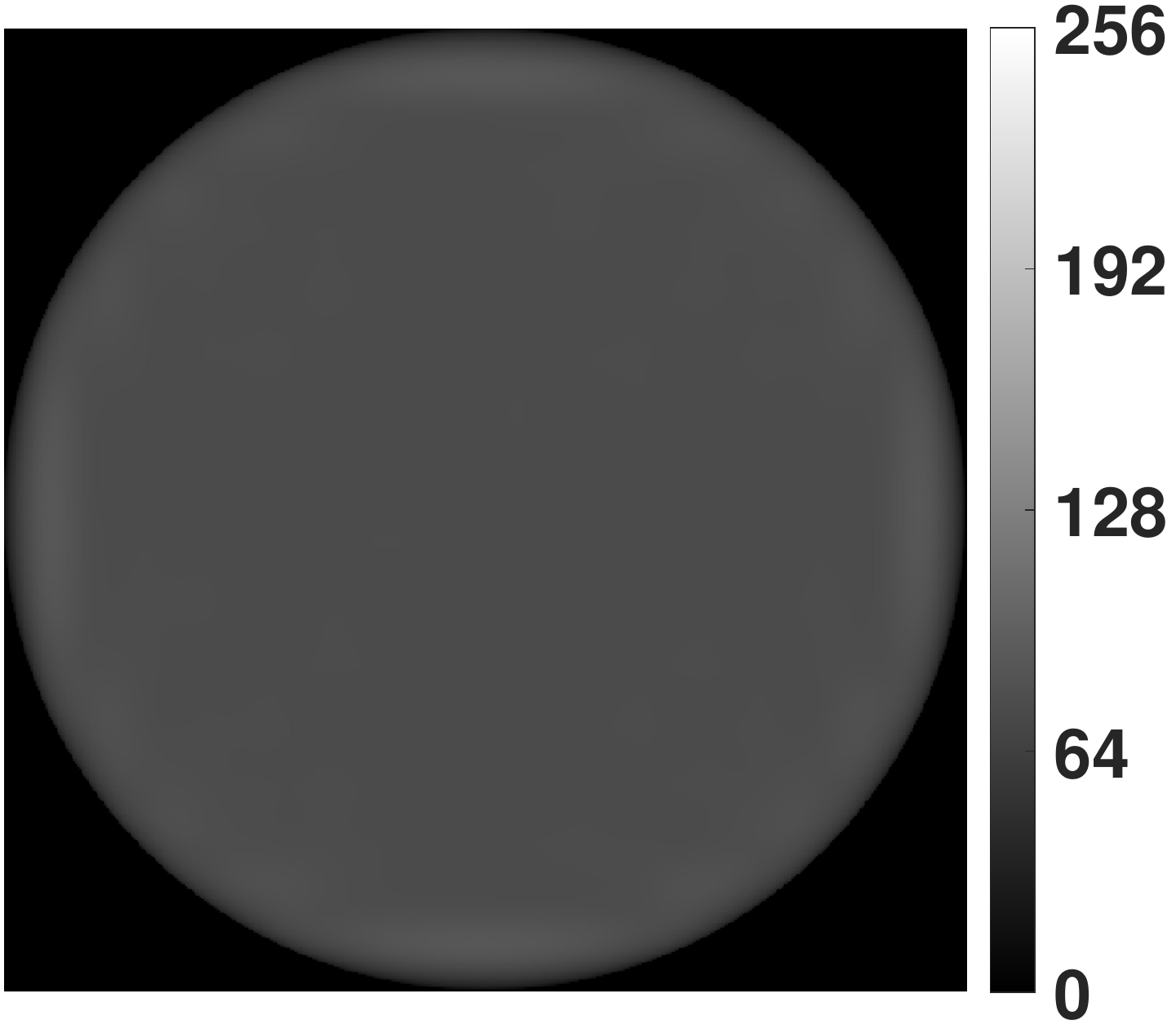}}\\
	
	\subfloat[\label{fig:principle-drr-breakout-sphere}]{\includegraphics[width=0.24\columnwidth]{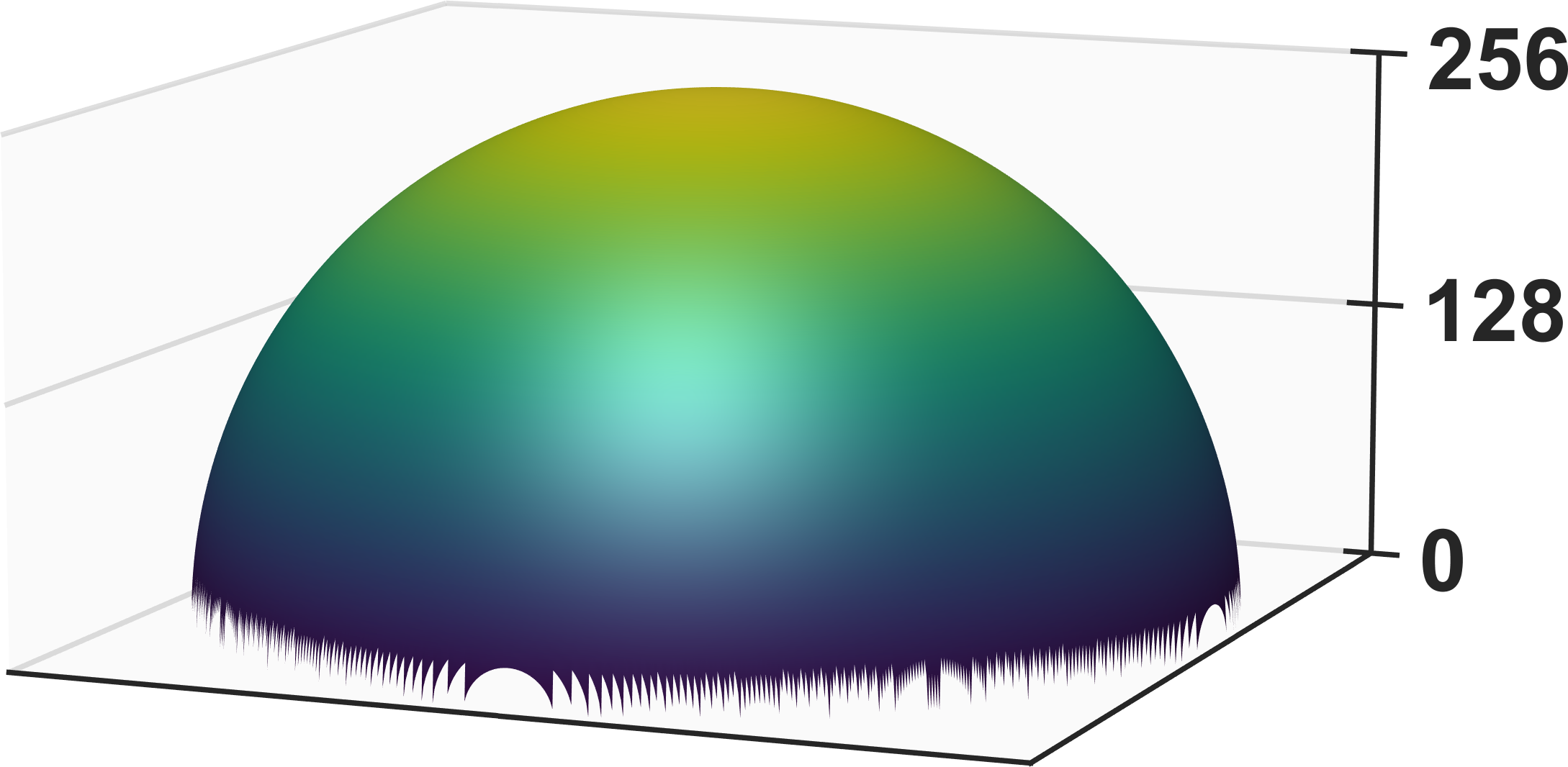}}
	\hspace{0.3em}
	\subfloat[\label{fig:principle-drr-breakout-sphere-approx}]{\includegraphics[width=0.24\columnwidth]{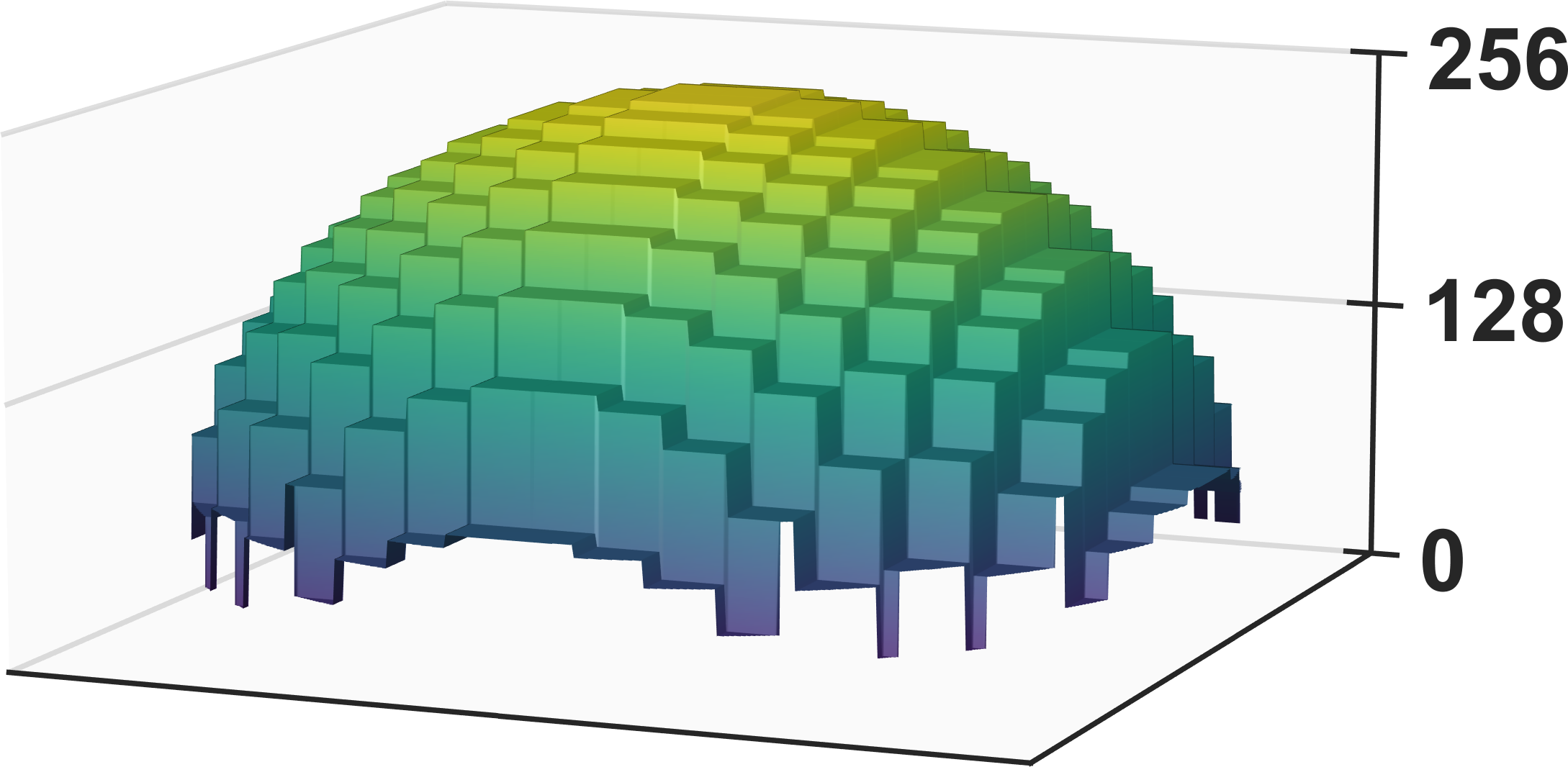}}
	\hspace{0.3em}
	\subfloat[\label{fig:principle-drr-breakout-sphere-approx-filtered}]{\includegraphics[width=0.24\columnwidth]{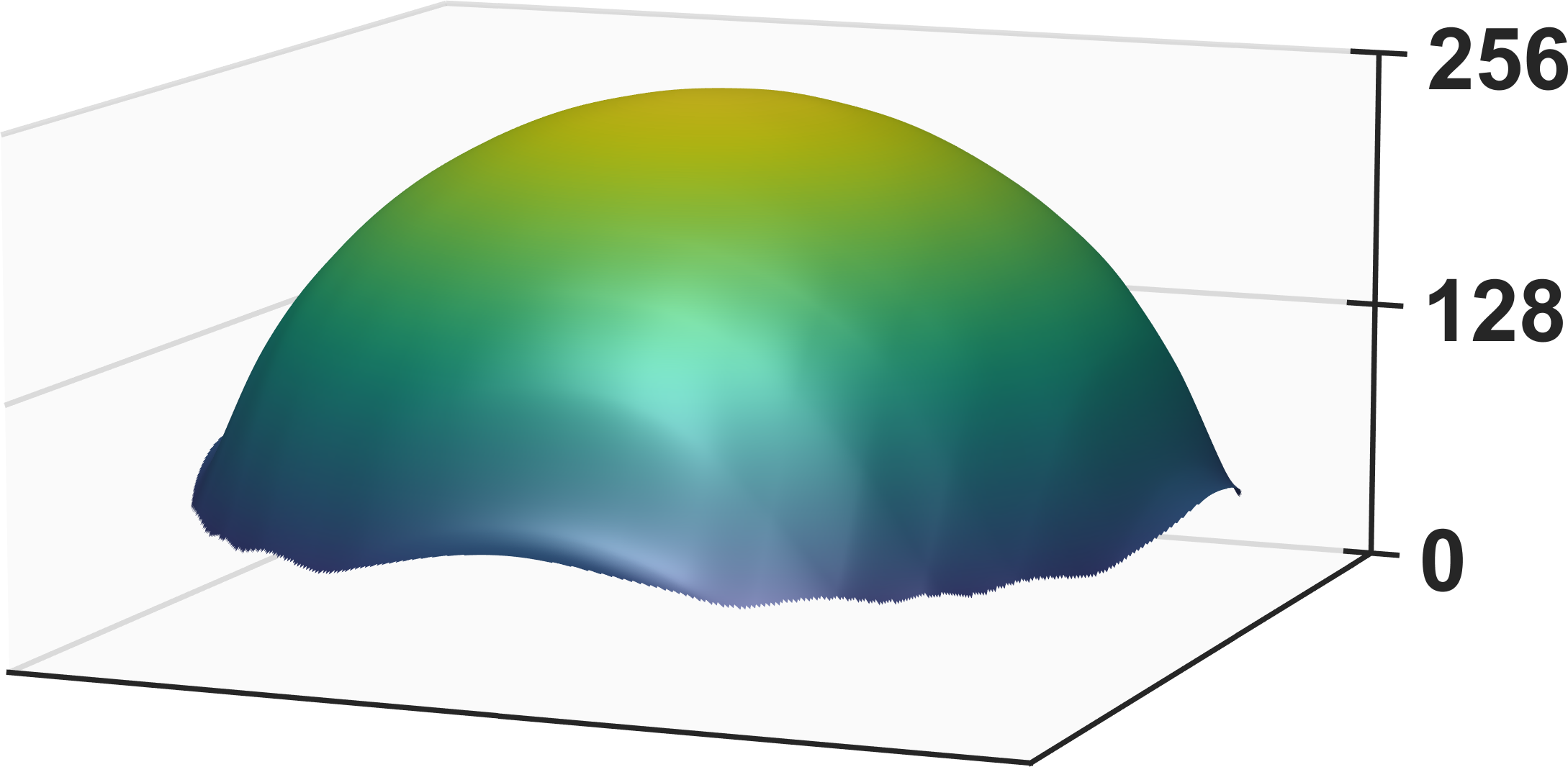}}
	\hspace{0.3em}
	\subfloat[\label{fig:principle-drr-breakout-sphere-reduced}]{\includegraphics[width=0.24\columnwidth]{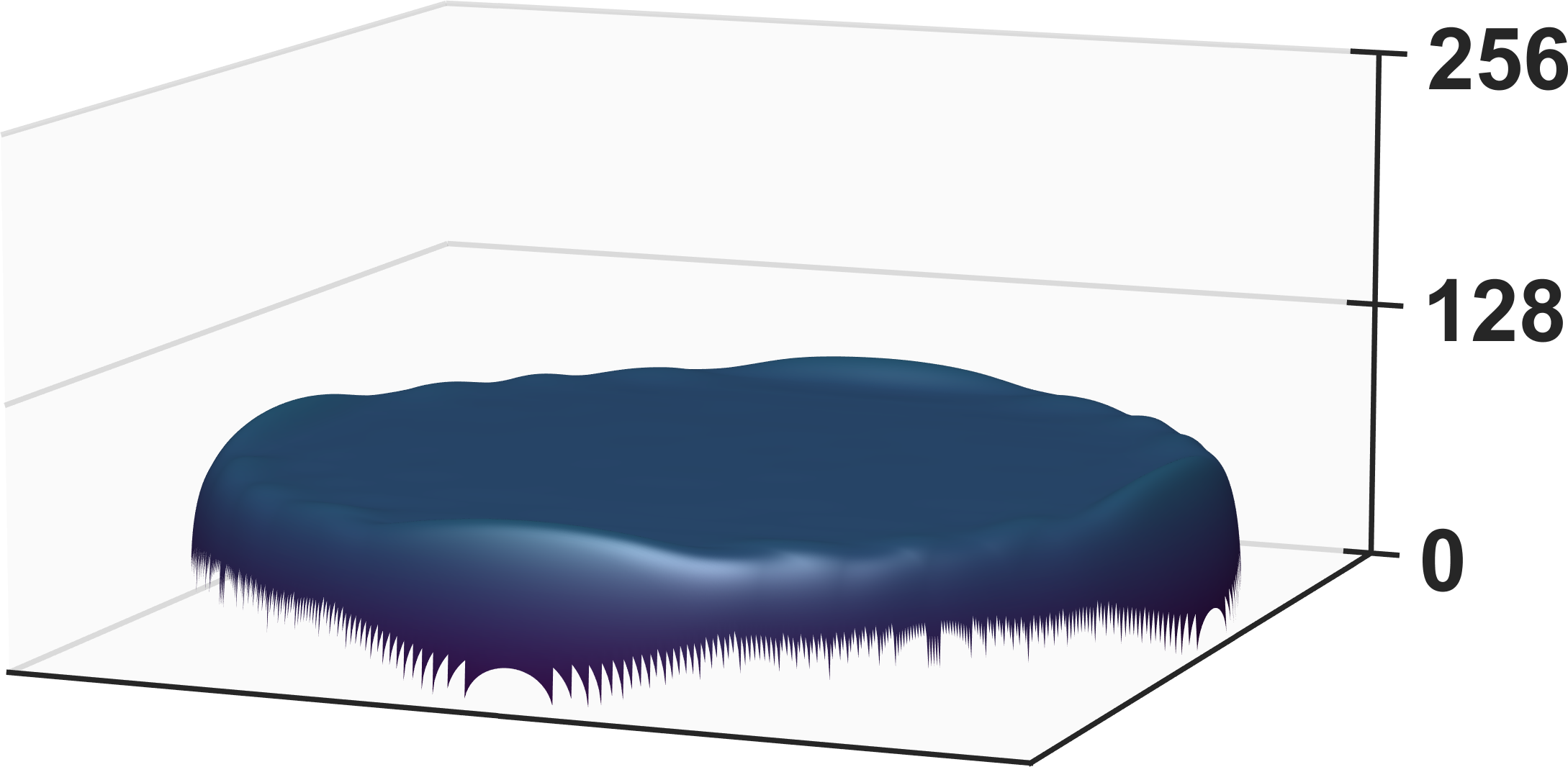}}\\
	
	\caption{
		The proposed method of depth range reduction applied to an ideal hemisphere with radius 256 mm. (a) the $512 \times 512$ 2D depth map to be encoded, $Z$; (b) a `blocked' approximation of $Z$, where each $32 \times 32$ neighborhood has been replaced by the mean value of the neighborhood being evaluated, leaving only $16 \times 16$ distinct pixel values; (c) the approximate geometry, $\widetilde{Z}'$, generated by resizing the $16 \times 16$ thumbnail image generated from (b); (d) the range-reduced depth map, $Z_r$, generated by subtracting (c) from (a); (e)-(h) 3D renderings corresponding to the depth maps (a)-(d), respectively.
	}
	
	\label{fig:principle-drr-breakout}
\end{figure}

Figure~\ref{fig:principle-drr-breakout} illustrates the proposed method applied to an ideal hemisphere with a radius and depth range of 256 mm. 
Figure~\ref{fig:principle-drr-breakout-depth} is $Z$, a $512 \times 512$ depth map representing the original 3D geometry.
Figure~\ref{fig:principle-drr-breakout-depth-approx} is a `blocked' approximation of $Z$, where each non-overlapping $32 \times 32$ neighborhood has been set to the mean value of the neighborhood.
Figure~\ref{fig:principle-drr-breakout-depth-approx} can be reduced to a $16 \times 16$ thumbnail image, as there are only $16 \times 16$ distinct pixel values present.
Figure~\ref{fig:principle-drr-breakout-depth-approx-filtered} is the approximated geometry, $\widetilde{Z}'$, which was generated by taking the $16 \times 16$ thumbnail approximation of $Z$ and using bicubic interpolation to scale the thumbnail image back to the original $512 \times 512$ image resolution via \emph{nanimresize}\cite{Greene:nanimresize:2020}.
Since $\widetilde{Z}$ was generated directly from the depth map $Z$, there is no need to perform an alignment step and $\widetilde{Z}' = \widetilde{Z}$.
Figure~\ref{fig:principle-drr-breakout-depth-reduced} is $Z_r$, the range-reduced depth map generated by subtracting Fig.~\ref{fig:principle-drr-breakout-depth-approx-filtered} from Fig.~\ref{fig:principle-drr-breakout-depth} via Eq.~\ref{eqn:drr:subtraction}. 
In this example, the depth range was reduced from 256 mm to 87.4 mm (65.9\% reduction in depth range). Figures~\ref{fig:principle-drr-breakout-sphere}-\ref{fig:principle-drr-breakout-sphere-reduced} are the equivalent 3D renderings of the depth maps found in Figs.~\ref{fig:principle-drr-breakout-depth}-\ref{fig:principle-drr-breakout-depth-reduced}. 

In summary, after an approximated and aligned depth map, $\widetilde{Z}'$, has been derived, the original depth map, $Z$, can have its depth range reduced via the subtraction in Eq.~(\ref{eqn:drr:subtraction}) to produce the range-reduced depth map, $Z_r$.
Next, a variety of depth-based 3D range geometry compression methods may be used to encode this new depth map into the color channels of a 2D image.
This 2D image can be further compressed with lossless (e.g., PNG) or lossy (e.g., JPEG) 2D image compression methods to achieve reduced file sizes.
Finally, this compressed 2D image, as well as any auxiliary information required to generate the approximated geometry (e.g., the $16 \times 16$ thumbnail image discussed in Fig.~\ref{fig:principle-drr-breakout}), can be stored or transmitted.

\subsection{Decoding with Depth Range Reduction}
\label{sub:principle-decoding}

The decoding process is the functional opposite of the operations performed in Sec.~\ref{sub:principle-encoding}.
First, the reduced range geometry, $Z_r$, is recovered from the encoded depth image output by a depth compression algorithm.
Next, the approximated geometry, $\widetilde{Z}'$, is recovered from the overhead information transmitted alongside the encoded depth map.
Finally, the original geometry, $Z$, can be recovered by summing the approximated geometry and the reduced range geometry. 
Mathematically, this operation can be defined as 
\begin{equation}
Z(i,j) = Z_r(i,j) + \widetilde{Z}'(i,j),
\label{eqn:drr:addition}
\end{equation}
where $Z$ is the original geometry being recovered, $Z_r$ is the reduced range geometry recovered from a depth compression algorithm, and $\widetilde{Z}'$ is the pixel-aligned depth approximation of $Z$.

\section{Experimental Results}
\label{sec:experiments}
\begin{figure}[h!]
	\centering
	\subfloat[\label{fig:experiments-drr-breakout-orgZRender}]{\includegraphics[width=0.23\columnwidth]{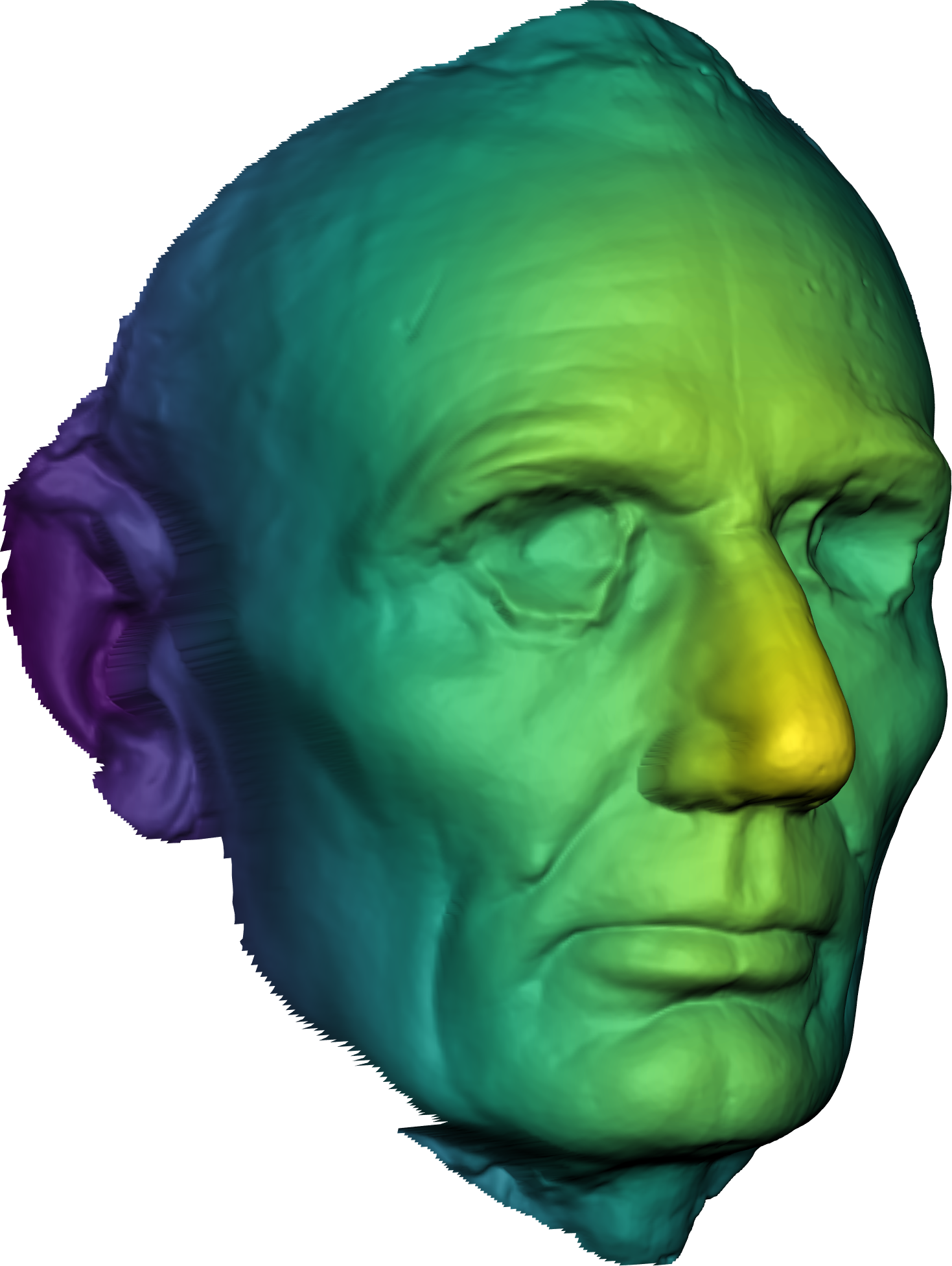}}
	\hspace{0.3em}
	\subfloat[\label{fig:experiments-drr-breakout-pixelImageRender}]{\includegraphics[width=0.23\columnwidth]{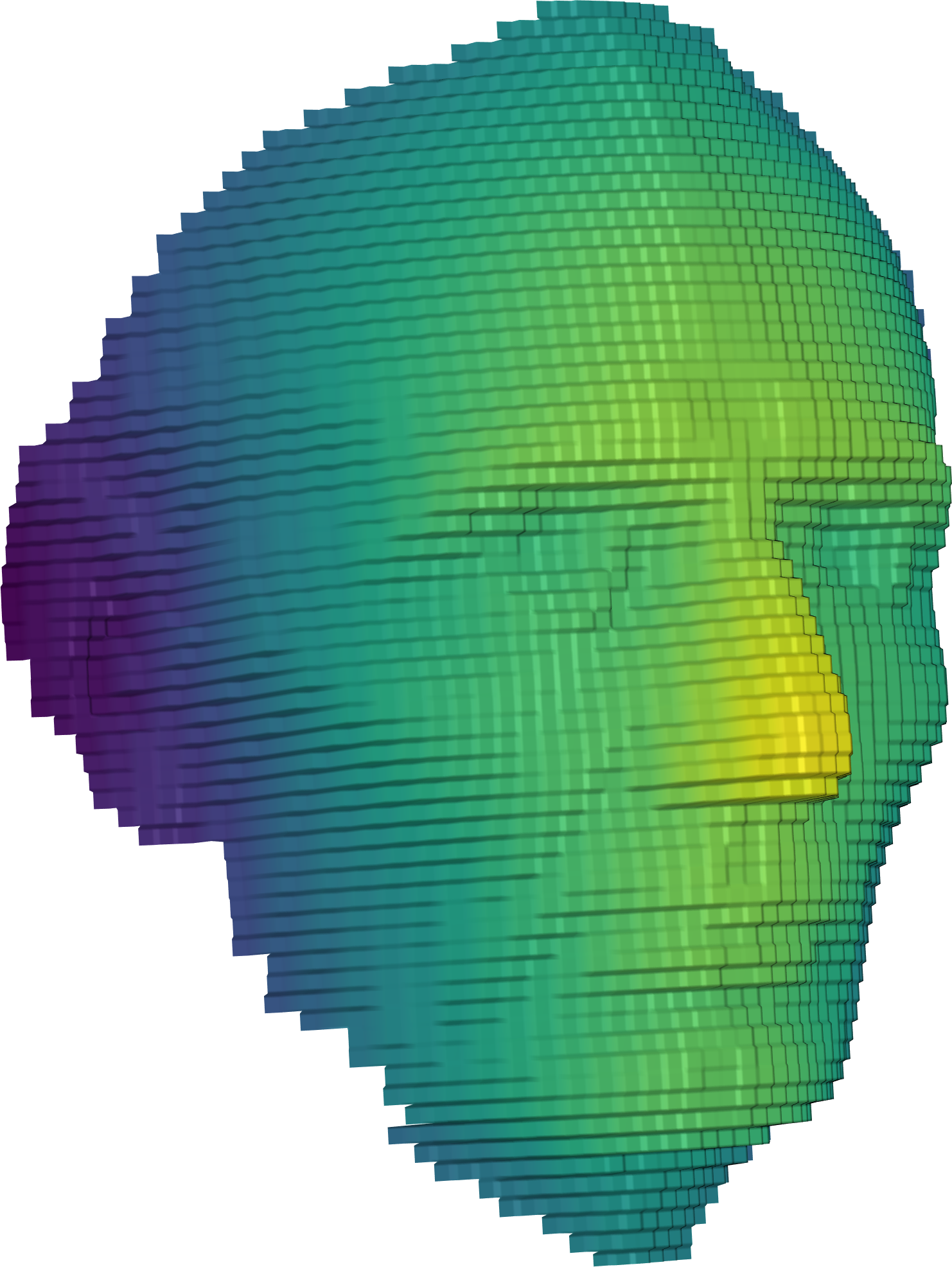}}
	\hspace{0.3em}
	\subfloat[\label{fig:experiments-drr-breakout-depthApproxRender}]{\includegraphics[width=0.23\columnwidth]{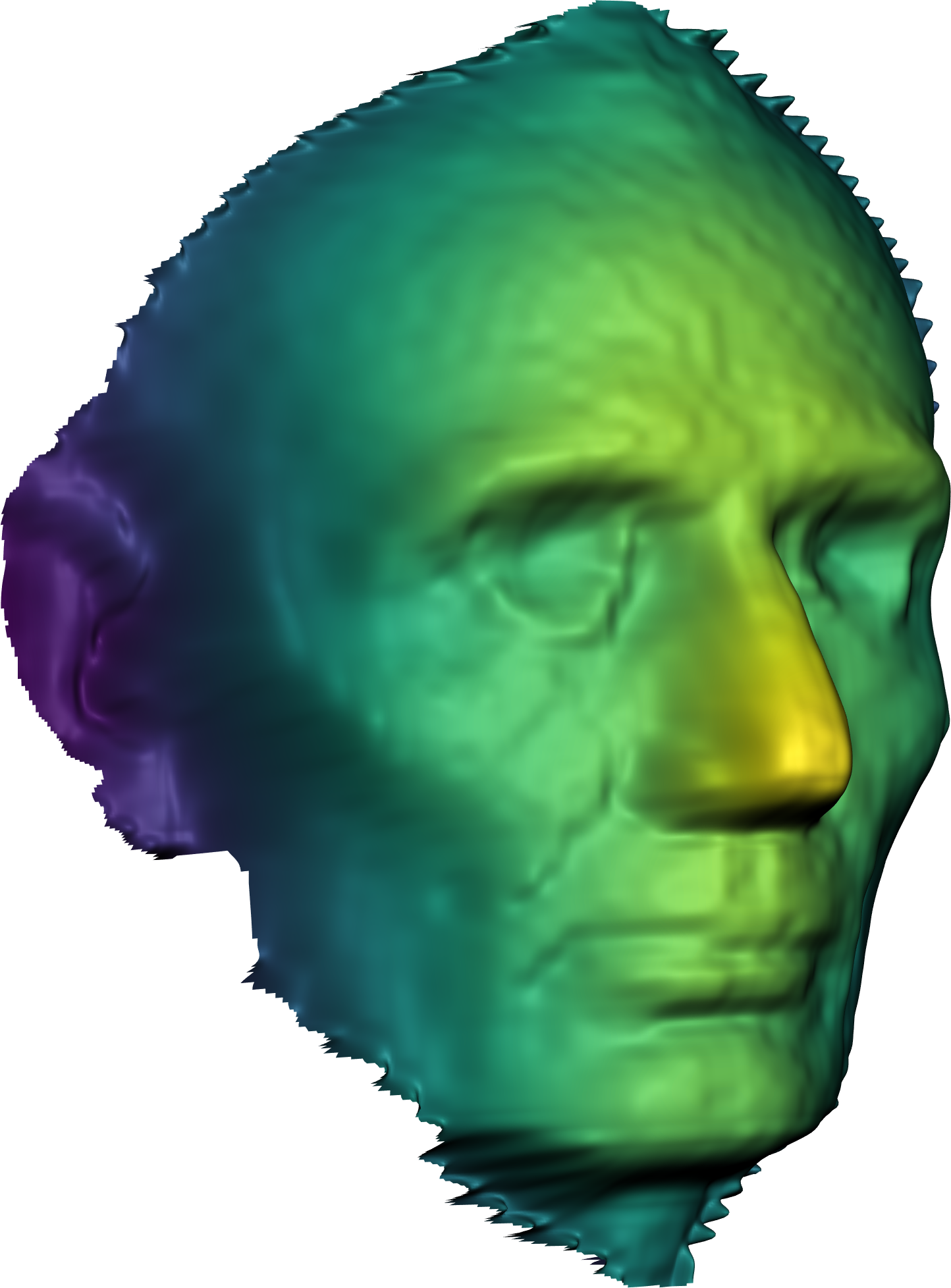}}
	\hspace{0.3em}
	\subfloat[\label{fig:experiments-drr-breakout-depthReducedRender}]{\includegraphics[width=0.23\columnwidth]{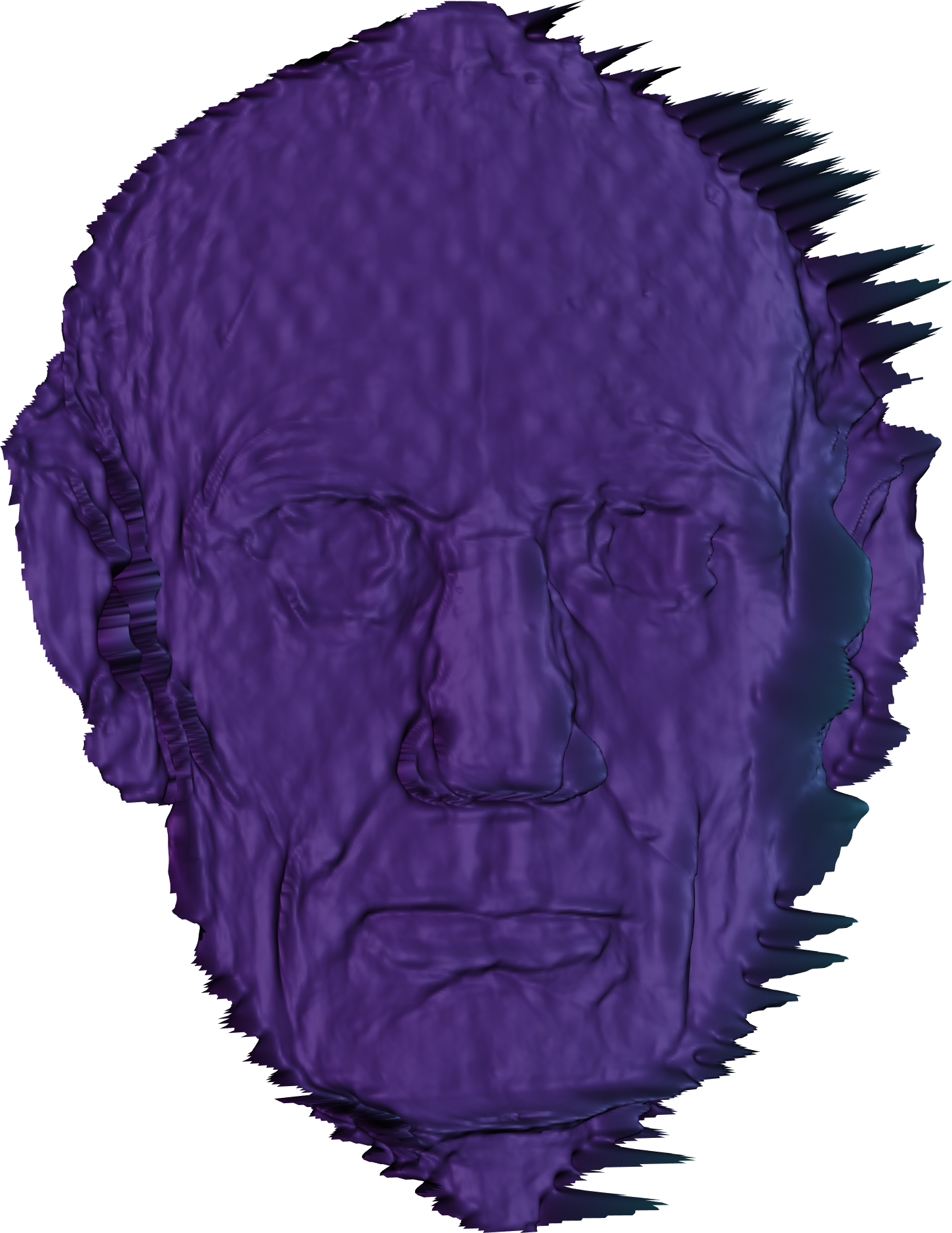}}\\
	
	\subfloat[\label{fig:experiments-drr-breakout-encodedImage}]{\includegraphics[width=0.23\columnwidth]{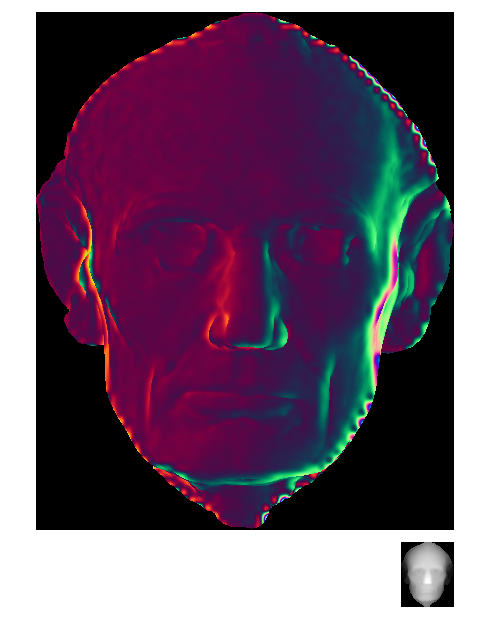}}
	\hspace{0.3em}
	\subfloat[\label{fig:experiments-drr-breakout-depthReducedRenderRecovered}]{\includegraphics[width=0.23\columnwidth]{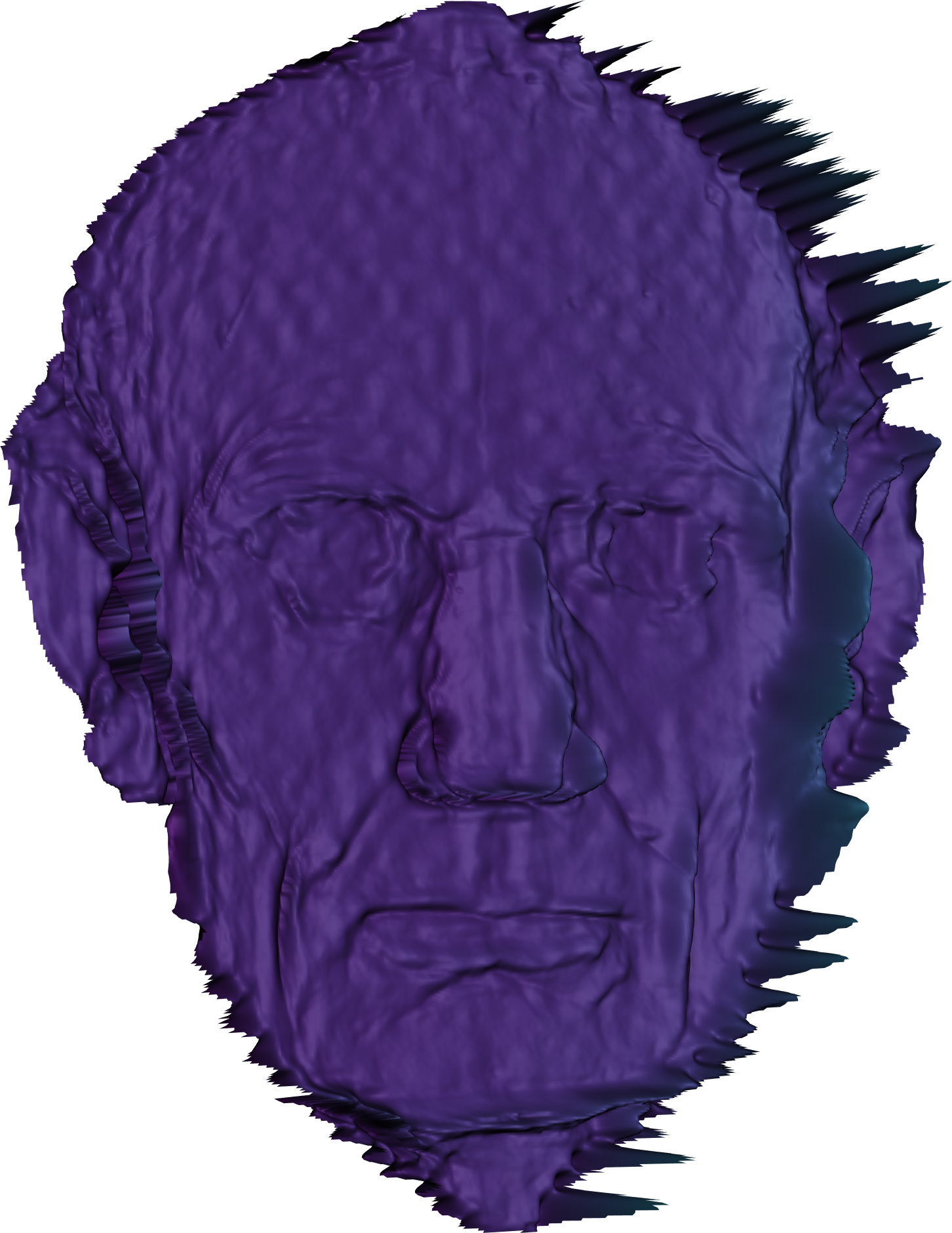}}
	\hspace{0.3em}
	\subfloat[\label{fig:experiments-drr-breakout-depthApproxRenderRecovered}]{\includegraphics[width=0.23\columnwidth]{figs/LincolnBreakout-DepthApprox-3D}}
	\hspace{0.3em}
	\subfloat[\label{fig:experiments-drr-breakout-orgZRenderRecovered}]{\includegraphics[width=0.23\columnwidth]{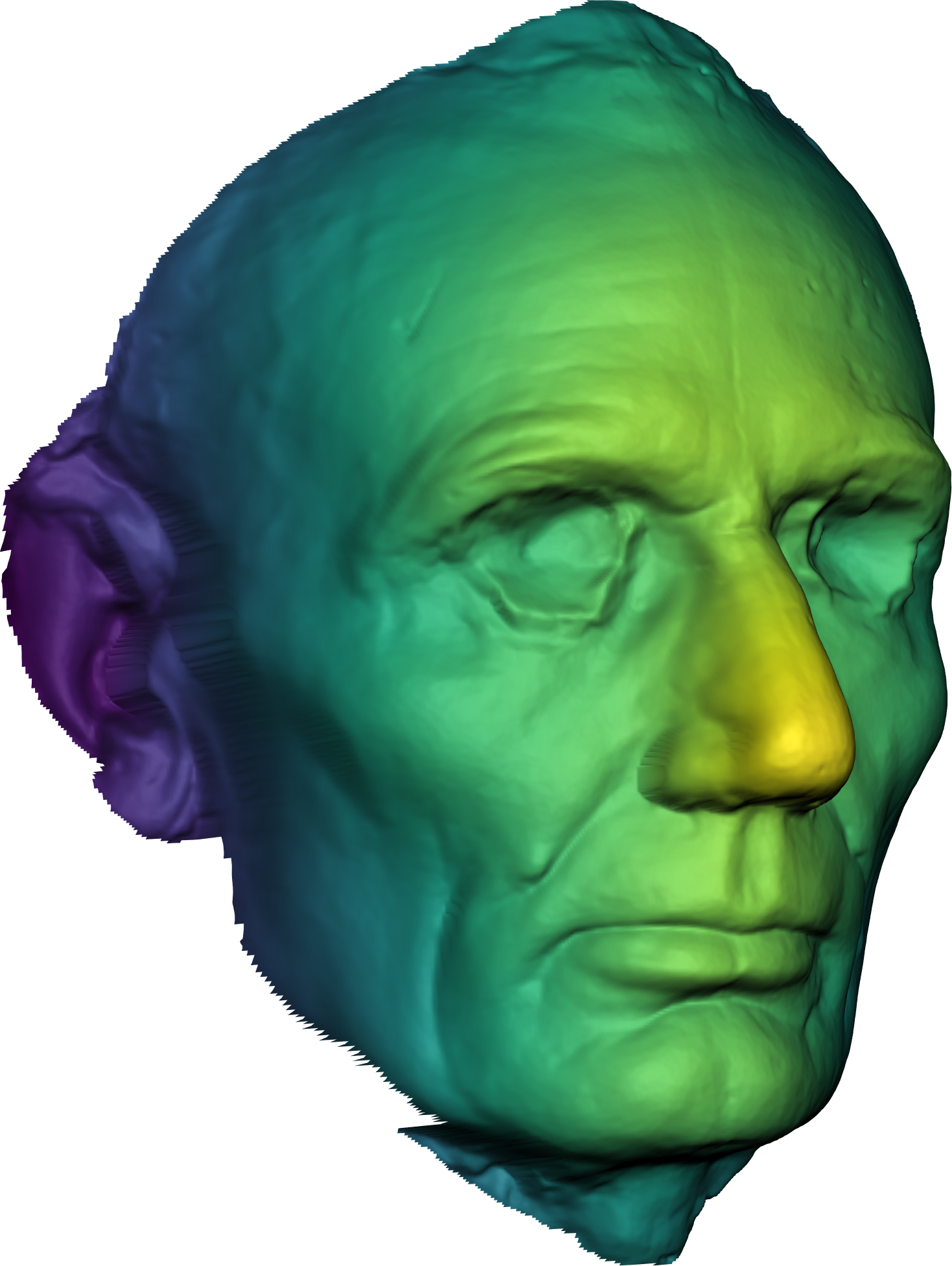}}\\
	
	\caption{The proposed method applied to a set of complex data. (a) 3D rendering of $Z$, the geometry to be encoded, with depth range 302.2 mm; (b) 3D rendering of the $65 \times 53$ thumbnail image used to approximate $Z$; (c) 3D rendering of $\widetilde{Z}'$, the approximation of $Z$ generated by resizing (b) to the original image dimensions; (d) 3D rendering of $Z_r$, the reduced range geometry, with depth range 128.5 mm, generated by subtracting (c) from (a); (e) the encoded image, stored in the PNG format, output from the MWD encoding method when $Z_r$ is used as the input with n = 2; the thumbnail image discussed in (b) can also be seen here, stored in the PNG format; (f) 3D rendering of $Z_r$, recovered using the MWD decoding process on the encoded image seen in (e); (g) 3D rendering of $\widetilde{Z}'$, recovered by resizing the thumbnail image seen in (e); (h) 3D rendering of $Z$, recovered by adding (g) to (f).}
	
	\label{fig:experiments-drr-breakout}
\end{figure}

The proposed depth range reduction method was evaluated using several experiments.
In the first experiment a piece of complex range geometry was used, in this case a range scan of a plaster cast of Abraham Lincoln's face~\cite{Smithsonian:LincolnModel}.
Figure~\ref{fig:experiments-drr-breakout-orgZRender} is a 3D rendering of $Z$, the original geometry to be encoded that has a depth range of 302.2 mm.
Figure~\ref{fig:experiments-drr-breakout-pixelImageRender} is a 3D rendering of the `blocked' representation of the original geometry, generated by setting every value in each distinct, non-overlapping $8 \times 8$ neighborhood to the mean value of the entire neighborhood.
This `blocked' representation was normalized and reduced to a $65 \times 53$ PNG thumbnail image, shown at the bottom right of Fig.~\ref{fig:experiments-drr-breakout-encodedImage}.
Figure~\ref{fig:experiments-drr-breakout-depthApproxRender} is a 3D rendering of the approximated geometry $\widetilde{Z}'$, generated by rescaling and resizing the thumbnail image to the original data's $518 \times 418$ image dimensions.
Figure~\ref{fig:experiments-drr-breakout-depthReducedRender} is a 3D rendering of the range-reduced geometry, $Z_r$, generated by subtracting $\widetilde{Z}'$ from $Z$ via Eq.~\ref{eqn:drr:subtraction}.
The depth range of $Z_r$ is 128.5 mm, 57.4\% less than the original depth range of 302.2 mm.
Figure~\ref{fig:experiments-drr-breakout-encodedImage} is the encoded output resulting from passing $Z_r$ into the multiwavelength depth (MWD) encoding method~\cite{Bell:MWD:2015} with two encoding periods ($n$ = 2), stored as a PNG image with file size 150.07 KB.
Included towards the bottom of Fig.~\ref{fig:experiments-drr-breakout-encodedImage} is the thumbnail image described previously, which is transmitted alongside the encoded image as overhead necessary to recover the original depth information.
In this case, the PNG thumbnail image had a file size which was 1.59 KB resulting in a total storage cost of 151.66 KB for the encoded data.
Figure~\ref{fig:experiments-drr-breakout-depthReducedRenderRecovered} is a 3D rendering of $Z_r$, recovered from Fig.~\ref{fig:experiments-drr-breakout-encodedImage} through the use of the MWD decoding process.
Figure~\ref{fig:experiments-drr-breakout-depthApproxRenderRecovered} is a 3D rendering of the depth approximation, $\widetilde{Z}'$, recovered by rescaling and resizing the overhead thumbnail image seen in Fig.~\ref{fig:experiments-drr-breakout-encodedImage} to the original image dimensions of $Z$.
Figure~\ref{fig:experiments-drr-breakout-orgZRenderRecovered} is a 3D rendering of the recovered geometry, $Z$, which was generated by adding the recovered $\widetilde{Z}'$ to the recovered $Z_r$ via Eq.~\ref{eqn:drr:addition}.
This recovered geometry has an RMS error of 0.0234 mm compared to the original geometry, resulting in 99.99\% reconstruction accuracy. 
This experiment highlights the ability of the proposed method to achieve a large level of depth reduction (57.4\% reduction in depth range compared to the original geometry) with a relatively low amount of overhead cost (1.59 KB) for complex pieces of geometry.

\begin{figure}[h!]
	\centering
	\subfloat[\label{fig:experiments-drr-plots-MWD_q0}]{\includegraphics[width=0.33\columnwidth]{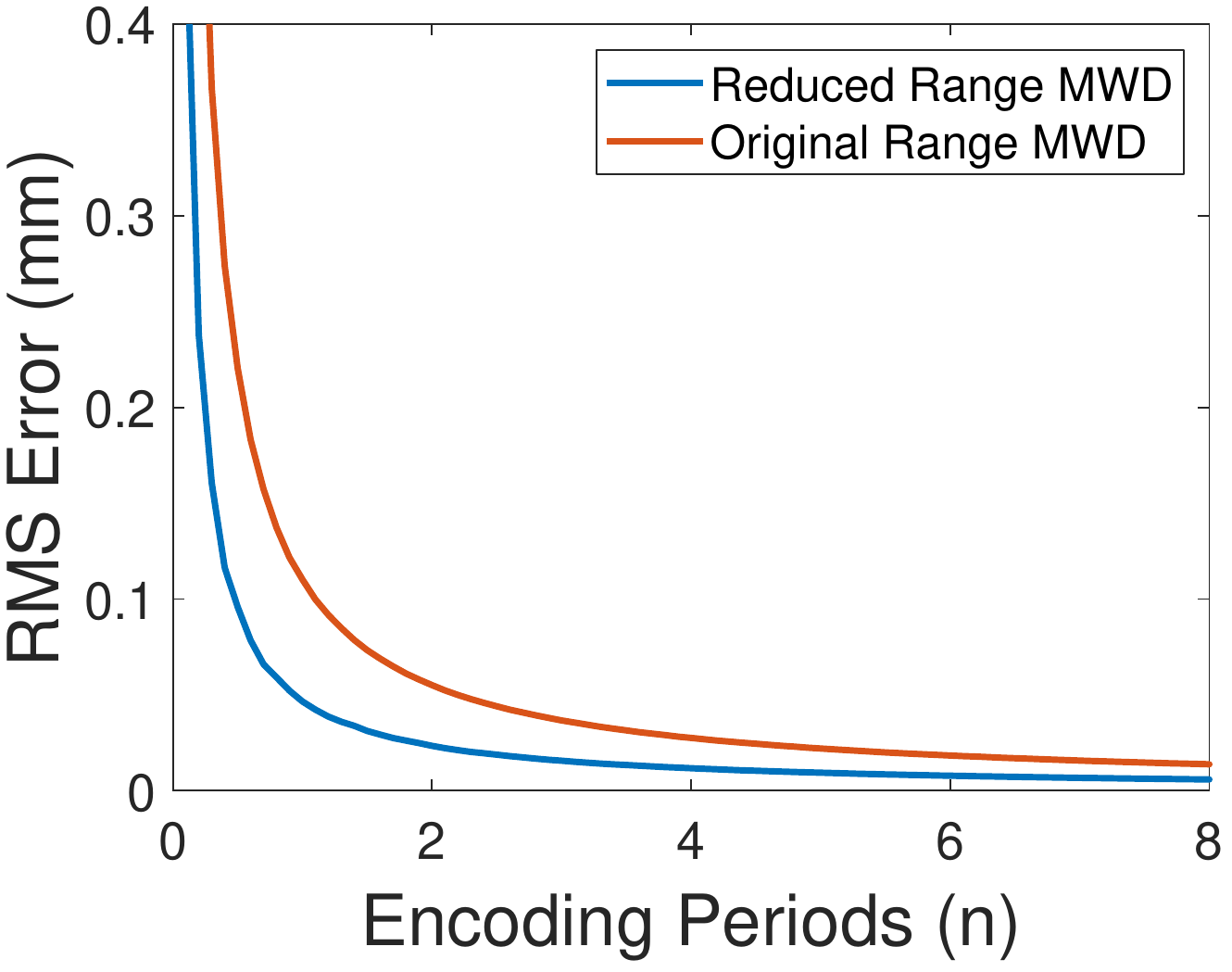}}
	\subfloat[\label{fig:experiments-drr-plots-MWD_q100}]{\includegraphics[width=0.33\columnwidth]{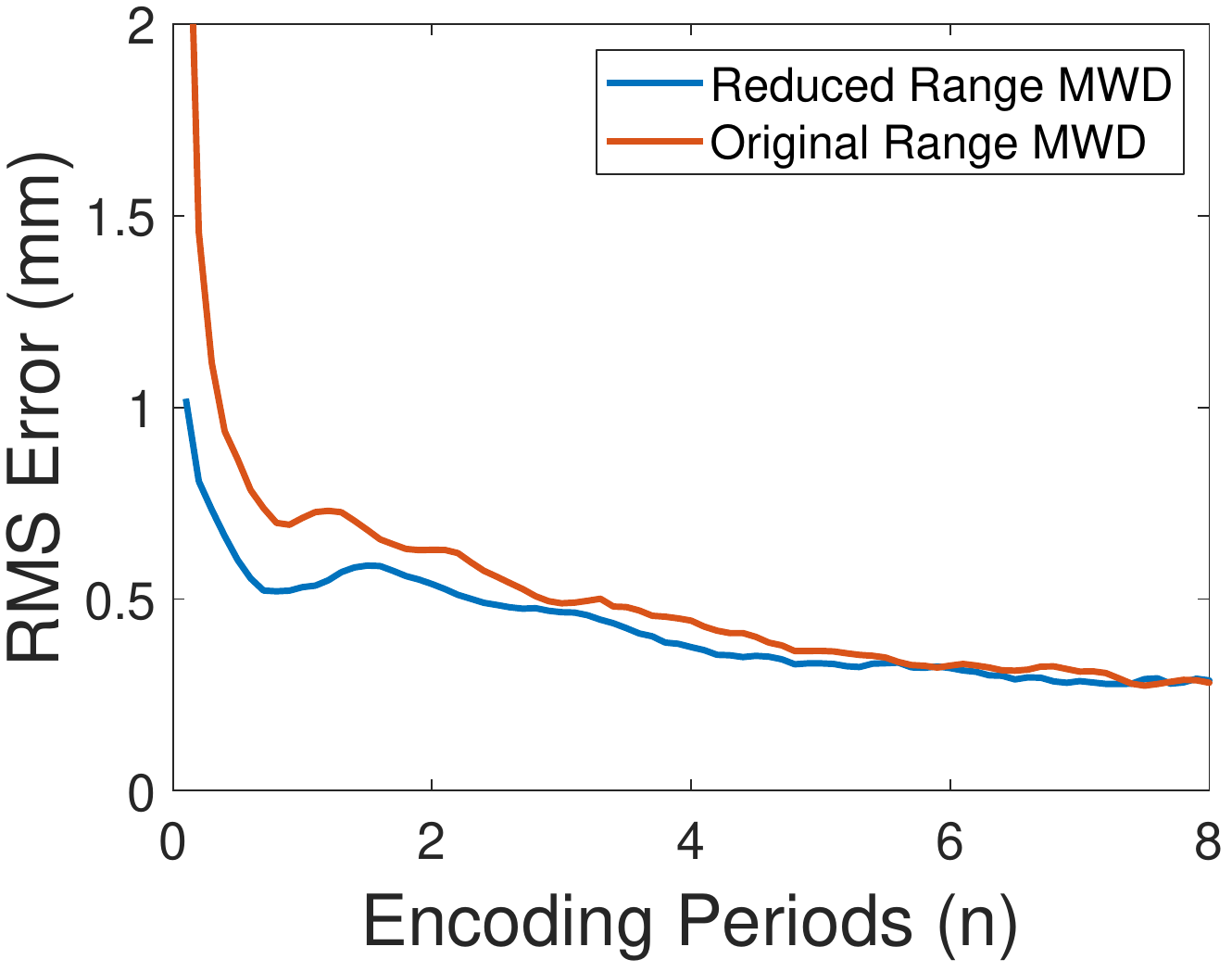}}
	\subfloat[\label{fig:experiments-drr-plots-MWD_q80}]{\includegraphics[width=0.33\columnwidth]{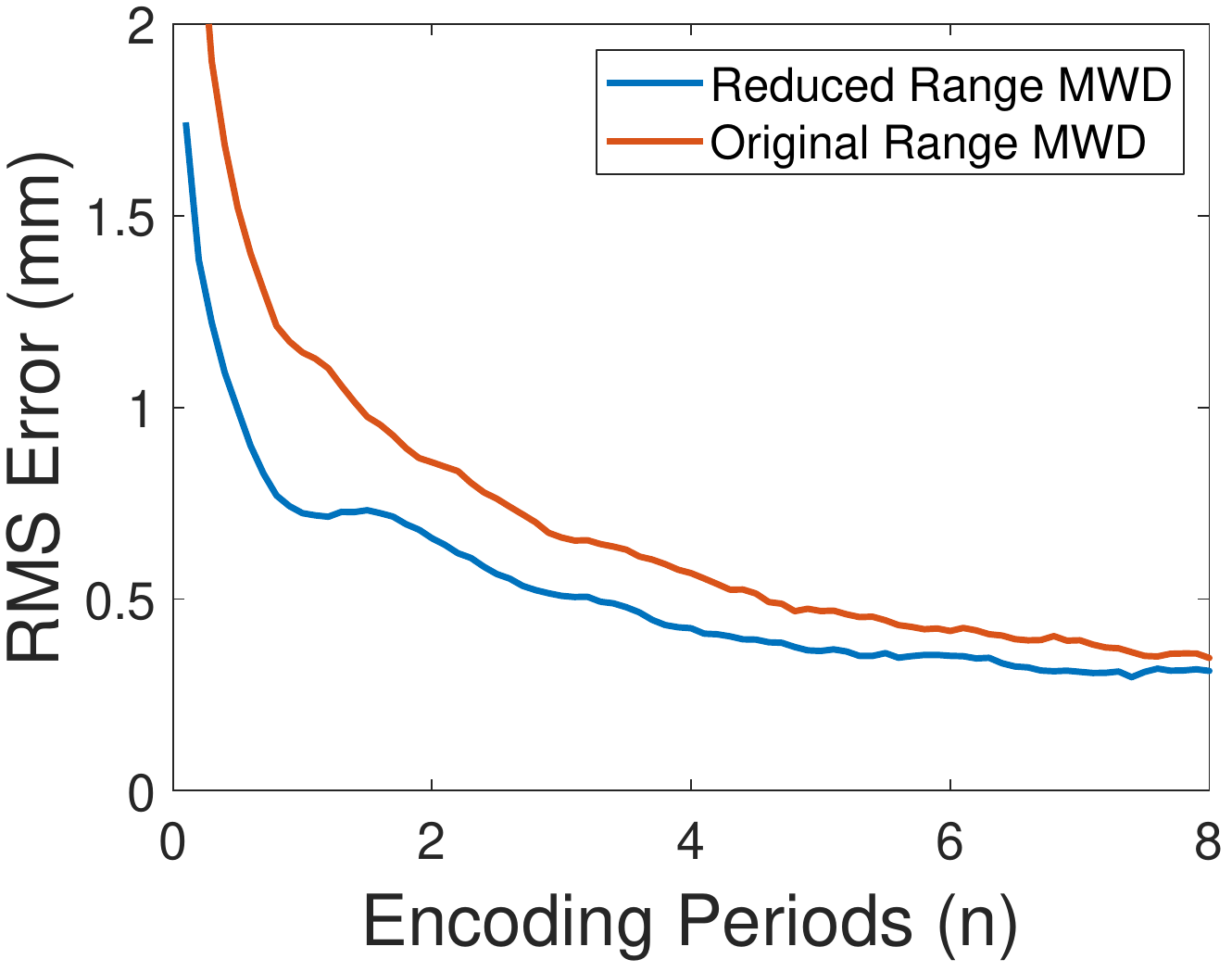}}\\
	
	\subfloat[\label{fig:experiments-drr-plots-DD_q0}]{\includegraphics[width=0.33\columnwidth]{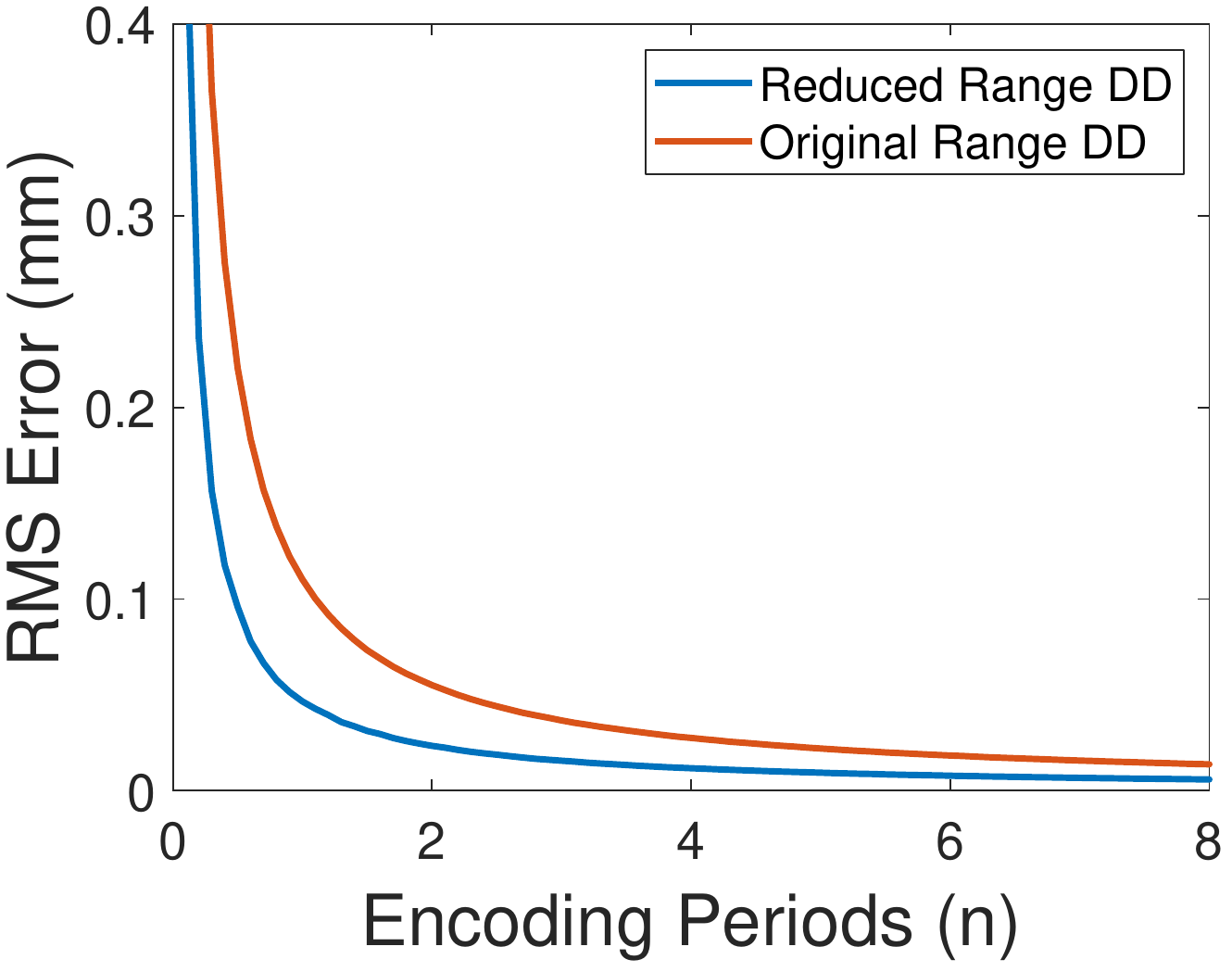}}
	\subfloat[\label{fig:experiments-drr-plots-DD_q100}]{\includegraphics[width=0.33\columnwidth]{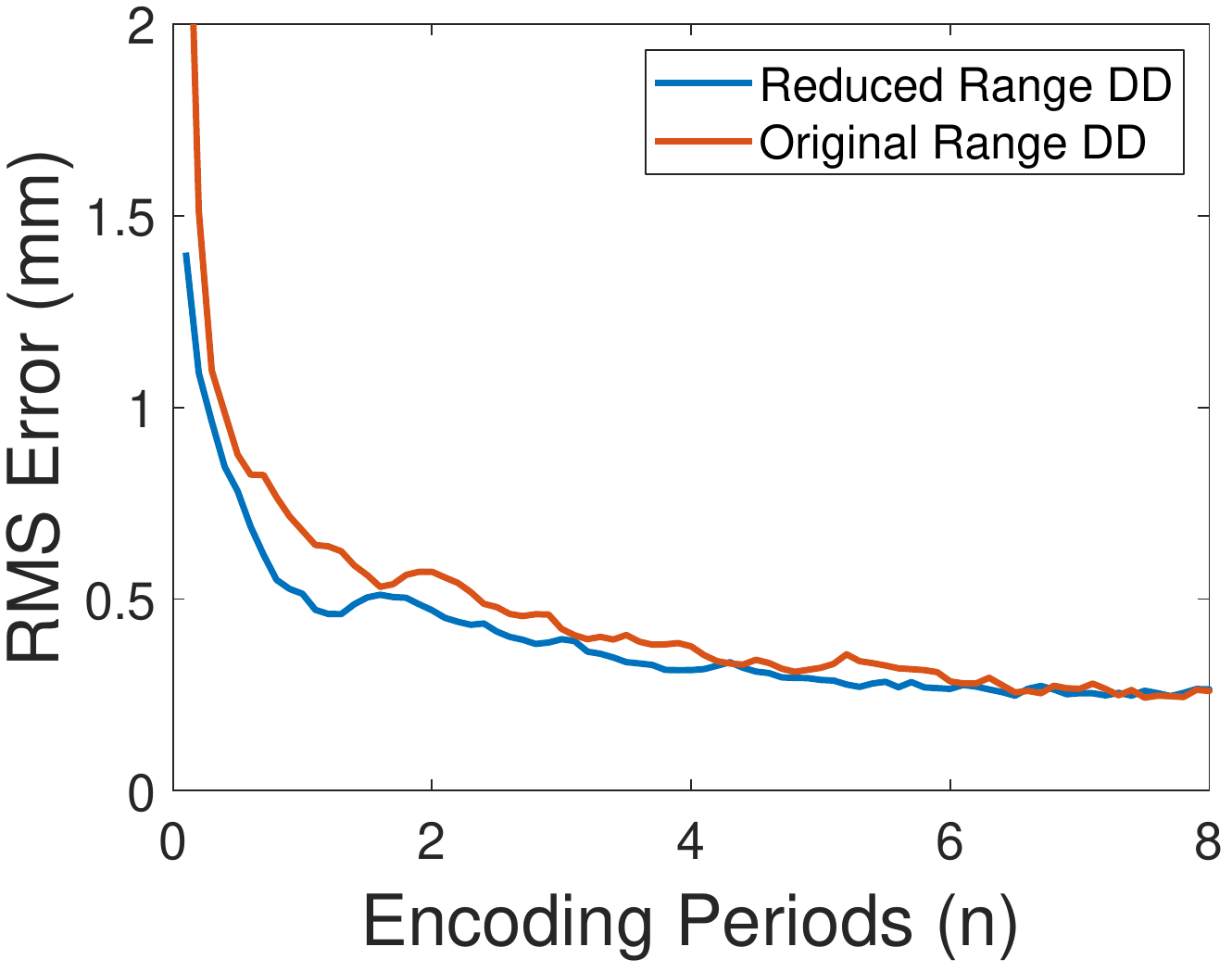}}
	\subfloat[\label{fig:experiments-drr-plots-DD_q80}]{\includegraphics[width=0.33\columnwidth]{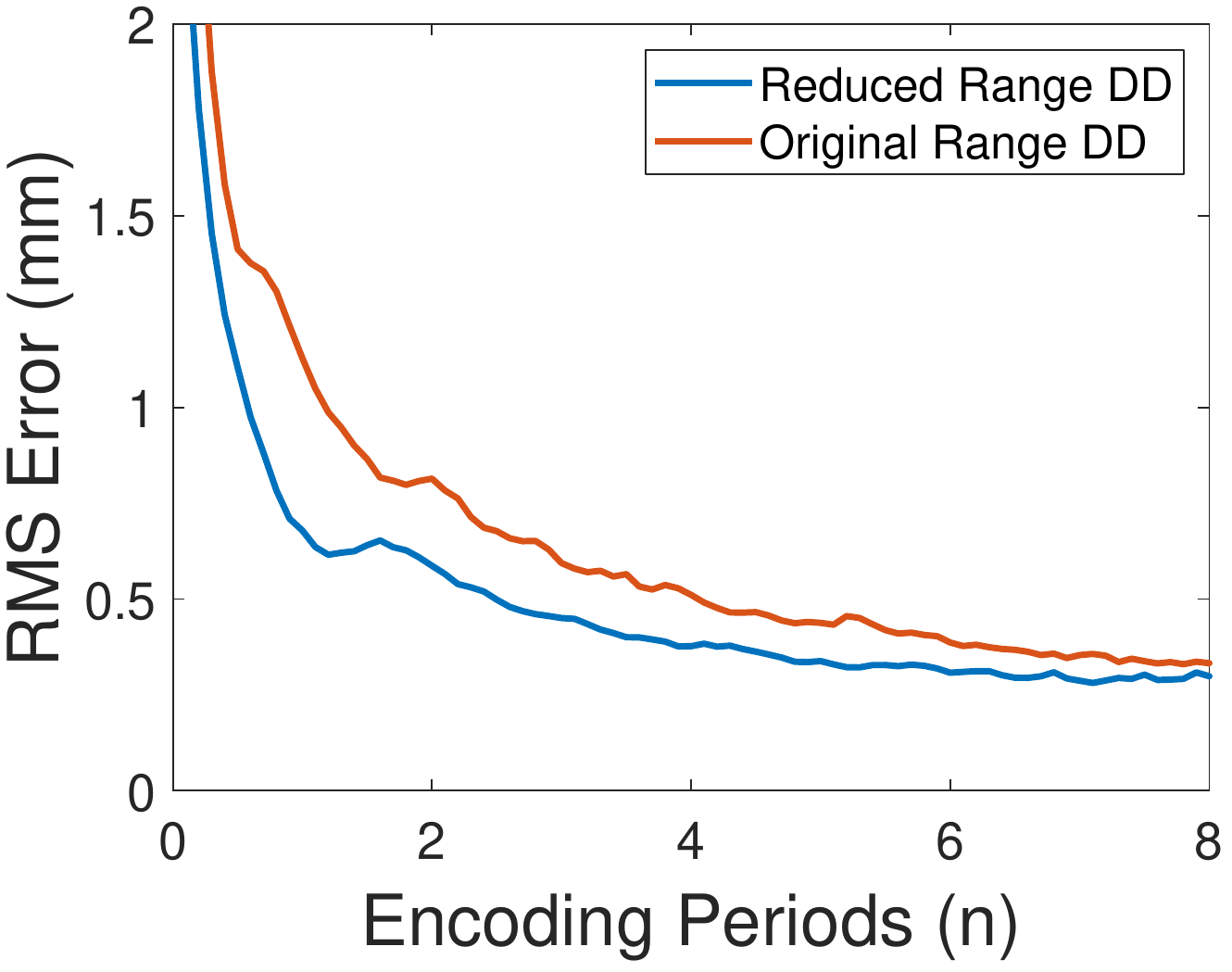}}\\
	
	\caption{The reconstruction error for two different depth compression methods with the proposed preprocessing method applied to the input compared with the reconstruction errors given the original input.
		(row 1) the RMS error between the recovered geometry and original geometry for the MWD encoding method when $Z$ is stored and when $Z_r$ is stored; (a)-(c) store their encoded image in the PNG format, the JPEG 100 format, and the JPEG 80 format, respectively.
		(row 2) the RMS error between the recovered geometry and original geometry for the DD encoding method when $Z$ is stored and when $Z_r$ is stored; (d)-(f) store their output image in the PNG format, the JPEG 100 format, and the JPEG 80 format, respectively.}
	
	\label{fig:experiments-drr-plots}
\end{figure}

The next experiment demonstrates that the proposed method of depth range reduction allows for a chosen minimum RMS reconstruction accuracy to be maintained while utilizing fewer encoding periods.
Figure~\ref{fig:experiments-drr-plots} shows the global reconstruction error values for the proposed method when storing the range-reduced geometry ($Z_r$) versus the error values when storing the original geometry ($Z$).
Additionally, these results are shown for two different range geometry compression algorithms: the MWD encoding method and the direct depth (DD) encoding method~\cite{Zhang:DirectDepth:2012}.
The same scan of a plaster cast of Abraham Lincoln's face and corresponding $65 \times 53$ approximation (1.59 KB file size) were used for this experiment as were used previously.
The first row gives results for the MWD encoding method, while the second row gives results for the DD encoding method.
The first column corresponds to the case where the output RGB image is stored in the PNG format, while the second and third columns use the JPEG format with quality set to 100 and 80, respectively.
To focus primarily on correctly analyzing unwrapped data, pixels with outlying absolute error values (greater than 10 mm) were ignored during the error analysis performed in this experiment.

For the PNG case, both the MWD and DD compression methods saw marked improvement when the depth range reduction preprocessing algorithm was applied: lower reconstruction error was achieved at any fixed number of encoding periods; conversely, fewer encoding periods were used to achieve any fixed target RMS error level.
The same trend can be seen when a lossy image storage method, in this case JPEG, was used to store the encoded $Z_r$: a smaller number of encoding periods were required to achieve a target reconstruction accuracy.
Given this observation it can be seen that the reduction of depth range in a set of complex 3D geometry allows it to be compressed---while maintaining some user-defined target precision---with fewer encoding periods.

\begin{figure}[h!]
	\centering
	\subfloat[\label{fig:experiments-drr-plots-annotated-error}]{\includegraphics[width=0.5\columnwidth]{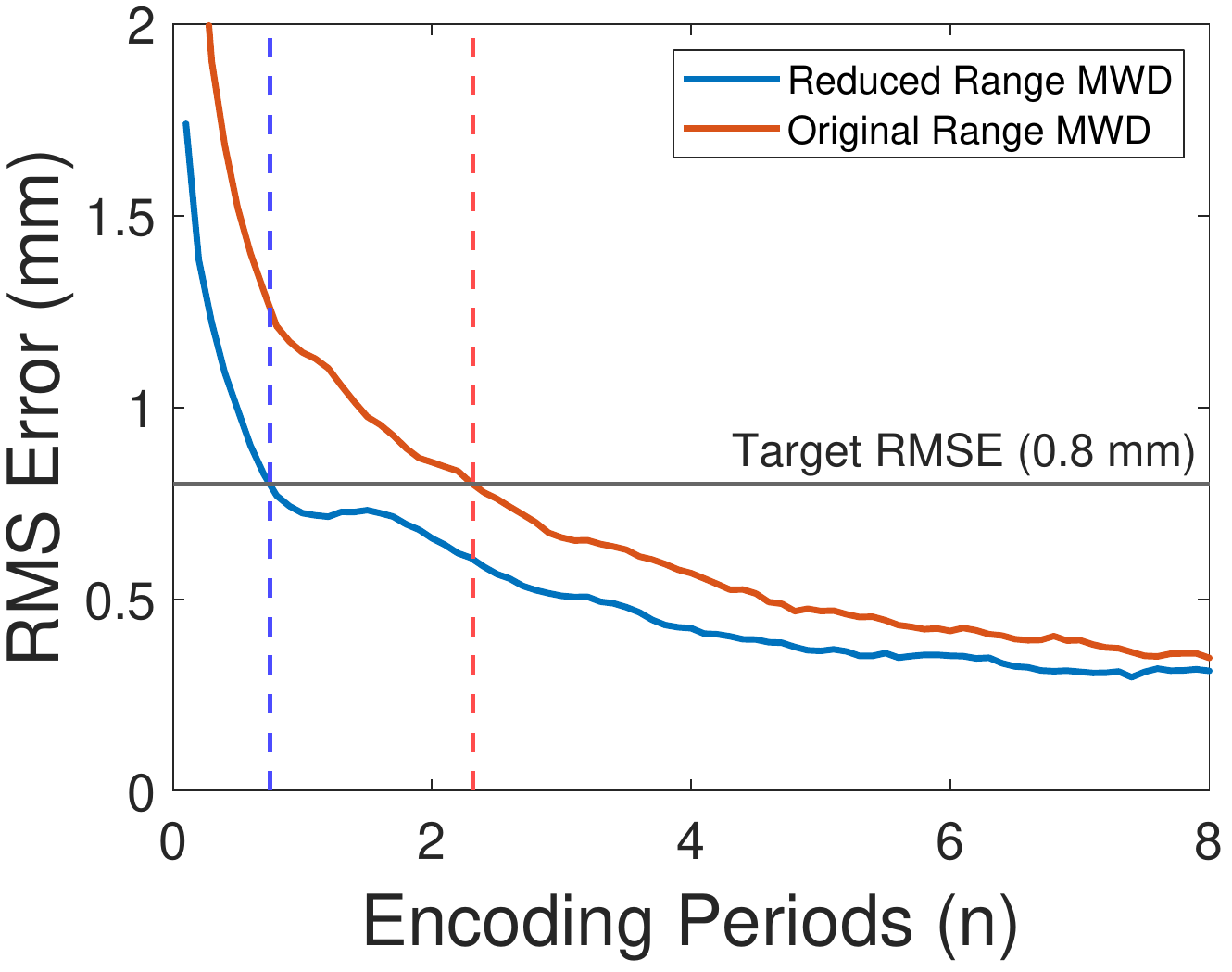}}
	\subfloat[\label{fig:experiments-drr-plots-annotated-fs}]{\includegraphics[width=0.5\columnwidth]{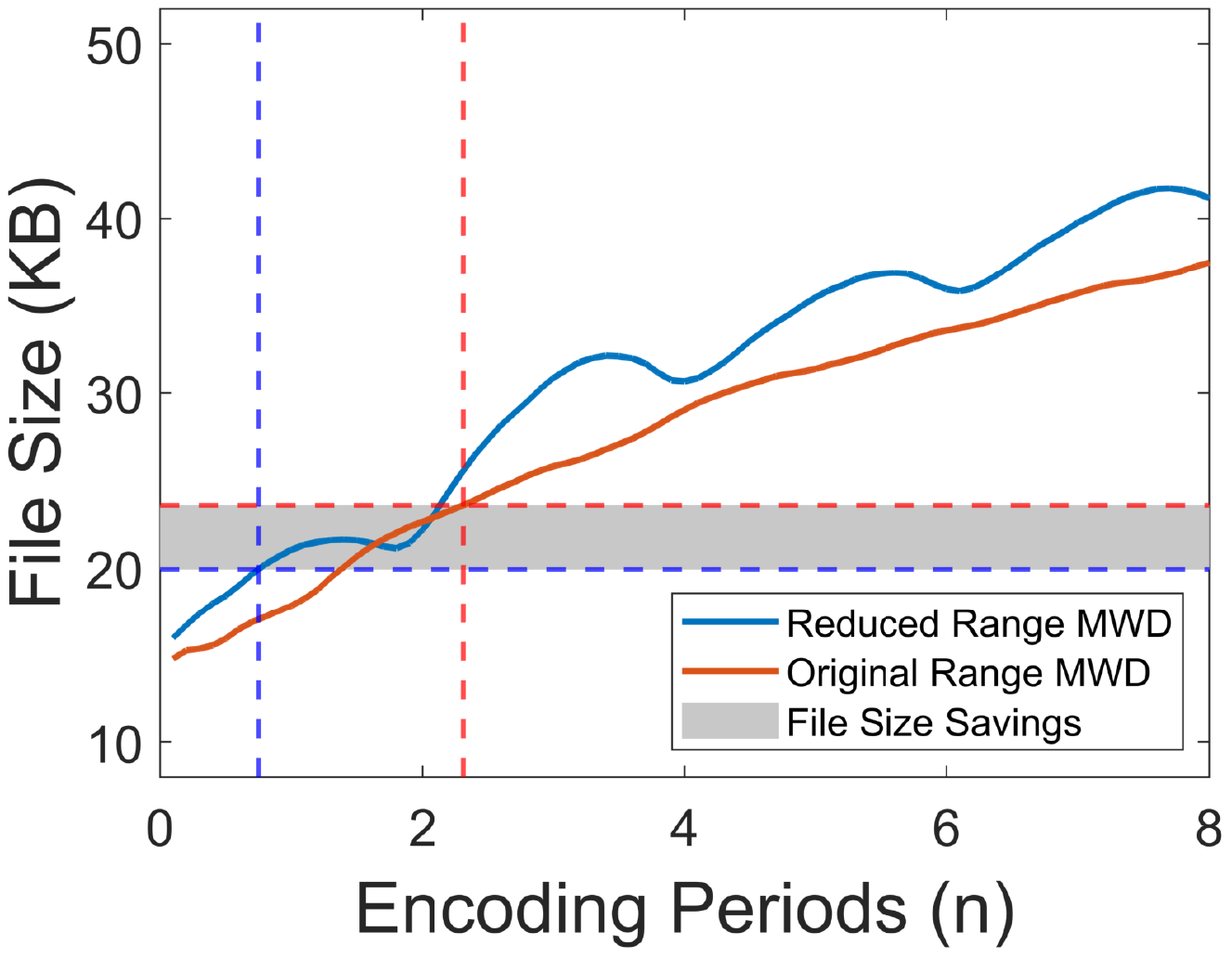}}\\
	
	\caption{Global reconstruction error values and associated file sizes for the MWD encoding method, with JPEG 80 as the image format used to store the encoded output, when $Z$ was used as the input compared to when $Z_r$ was used as the input. (a) the RMS error between the recovered and original geometry when $Z$ was used as input and when $Z_r$ was used as input; (b) the file size of MWD-encoded output image when $Z$ was used as input and when $Z_r$ was used as input.}
	
	\label{fig:experiments-drr-plots-annotated}
\end{figure}

The next experiment illustrates the ability of the proposed depth range reduction method to reduce file sizes while meeting user-defined constraints on maximum tolerable reconstruction error.
The same scan of a plaster cast of Abraham Lincoln's face, with original range 302.2 mm, was used in this experiment as was used previously.
Additionally, the same $65 \times 53$, 1.59 KB thumbnail image was used in this experiment to generate $\widetilde{Z}'$, resulting in a reduced depth range of 128.5 mm via the subtraction in Eq.~\ref{eqn:drr:subtraction}.
It should be noted that the thumbnail's file size is included anywhere the range-reduced encoded file size is mentioned in this experiment.
Figure~\ref{fig:experiments-drr-plots-annotated-error} shows the reconstruction error values for the MWD encoding method with the range-reduced geometry, $Z_r$, as input and the error values for the MWD encoding method with the original geometry, $Z$, as input.
This experiment used JPEG 80 in order to store the MWD-encoded output images.
The horizontal line in Fig.~\ref{fig:experiments-drr-plots-annotated-error} corresponds to some user-defined target reconstruction error; in this case 0.8 mm was chosen.
The red and blue dashed vertical lines in Fig.~\ref{fig:experiments-drr-plots-annotated-error} correspond to the smallest number of encoding periods needed to meet the target error requirement when using the MWD method to store $Z$ and $Z_r$, respectively.
Figure~\ref{fig:experiments-drr-plots-annotated-fs} shows the file sizes associated with the JPEG 80 encoded images output by the MWD method with the reduced range input and the original input.
The red and blue dashed vertical lines in Fig.~\ref{fig:experiments-drr-plots-annotated-fs} are identical to the dashed lines in Fig.~\ref{fig:experiments-drr-plots-annotated-error}.
The red and blue dashed horizontal lines in Fig.~\ref{fig:experiments-drr-plots-annotated-fs} correspond to the file sizes generated by MWD when using the required number of encoding periods to reach the target reconstruction error value for both the original and range-reduced input, respectively.
The gray shaded area in Fig.~\ref{fig:experiments-drr-plots-annotated-fs} highlights the file size savings attained when MWD was used with the proposed range reduction method applied to the input.
When 0.8 mm RMS reconstruction error was targeted, the file size was reduced from 23.58 KB to 19.92 KB (15.5\% reduction) when the proposed method of depth range reduction was applied to the input.

Table~\ref{tab:file-size-reduction} further highlights the proposed method's ability to achieve file size savings (for a variety of output image storage formats) while reaching a target minimum reconstruction accuracy. 
Table~\ref{tab:file-size-reduction} compares the number of encoding periods (and corresponding file sizes) required by the MWD encoding method to reach a fixed target RMS error value for both the original range geometry and the reduced range geometry. 
The range geometry, and the method of approximation used to reduce the depth range, is the same as discussed in the previous experiment.
However, this experiment varies the 2D image compression format used to store the MWD-encoded output image.
Table~\ref{tab:file-size-reduction-PNG} compares the proposed method of depth range reduction to the unaltered method when the PNG format is used to store the encoded output image.
Table~\ref{tab:file-size-reduction-JPG100} gives the comparison of when the JPEG format with quality 100 is used to store the encoded output image.
Table~\ref{tab:file-size-reduction-JPG80} gives the comparison of when the JPEG format with quality 80 is used to store the encoded output image.
The file size of the thumbnail image used by the proposed method, saved in the PNG format throughout this experiment (1.59 kilobytes), is included in the calculation for file size where applicable.
The proposed method of depth range reduction allows for reduced file sizes while maintaining some target level of reconstruction accuracy.
Additionally, this method is shown to work for both lossless and lossy 2D image compression methods.

\begin{table}[h!]
	\caption{Performance of the proposed method applied to MWD compared to MWD with the original depth range when compressing an ideal sphere. (a) comparison when PNG was used to store the encoded output image; (b) comparison when JPEG 100 was used to store the encoded output image; (c) comparison when JPEG 80 was used to store the encoded output image.}
	\centering
	\subfloat[]{
		\begin{tabular}{|l|c|c|c|}
			\hline
			\textbf{PNG}      & RMS Error  & Encoding Periods & File Size  \\ \hline
			Original Range, $Z$     & 0.1 mm       & 1.099          &  108.76 KB         \\ \hline
			Reduced Range, $Z_r$		  & 0.1 mm       & 0.481       &  99.15 KB         \\ \hline
		\end{tabular}
		\label{tab:file-size-reduction-PNG}
	}
	
	\subfloat[]{
		\begin{tabular}{|l|c|c|c|}
			\hline
			\textbf{JPEG 100}  & RMS Error       & Encoding Periods & File Size   \\ \hline
			Original Range, $Z$    & 0.6 mm        & 2.284           &  94.26 KB    \\ \hline
			Reduced Range, $Z_r$	  & 0.6 mm        & 0.506       & 68.85 KB    \\ \hline
		\end{tabular}
		\label{tab:file-size-reduction-JPG100}
	}
	
	\subfloat[]{
		\begin{tabular}{|l|c|c|c|}
			\hline
			\textbf{JPEG 80}  & RMS Error        & Encoding Periods & File Size  \\ \hline
			Original Range, $Z$     & 0.8 mm        & 2.315          &  23.58 KB    \\ \hline
			Reduced Range, $Z_r$		  & 0.8 mm        & 0.749       &  19.92 KB    \\ \hline
		\end{tabular}
		\label{tab:file-size-reduction-JPG80}
	}
	\label{tab:file-size-reduction}
\end{table}

\begin{figure}[t!]
	\centering
	\subfloat[\label{fig:experiments-apple-depth}]{\includegraphics[width=0.16\columnwidth]{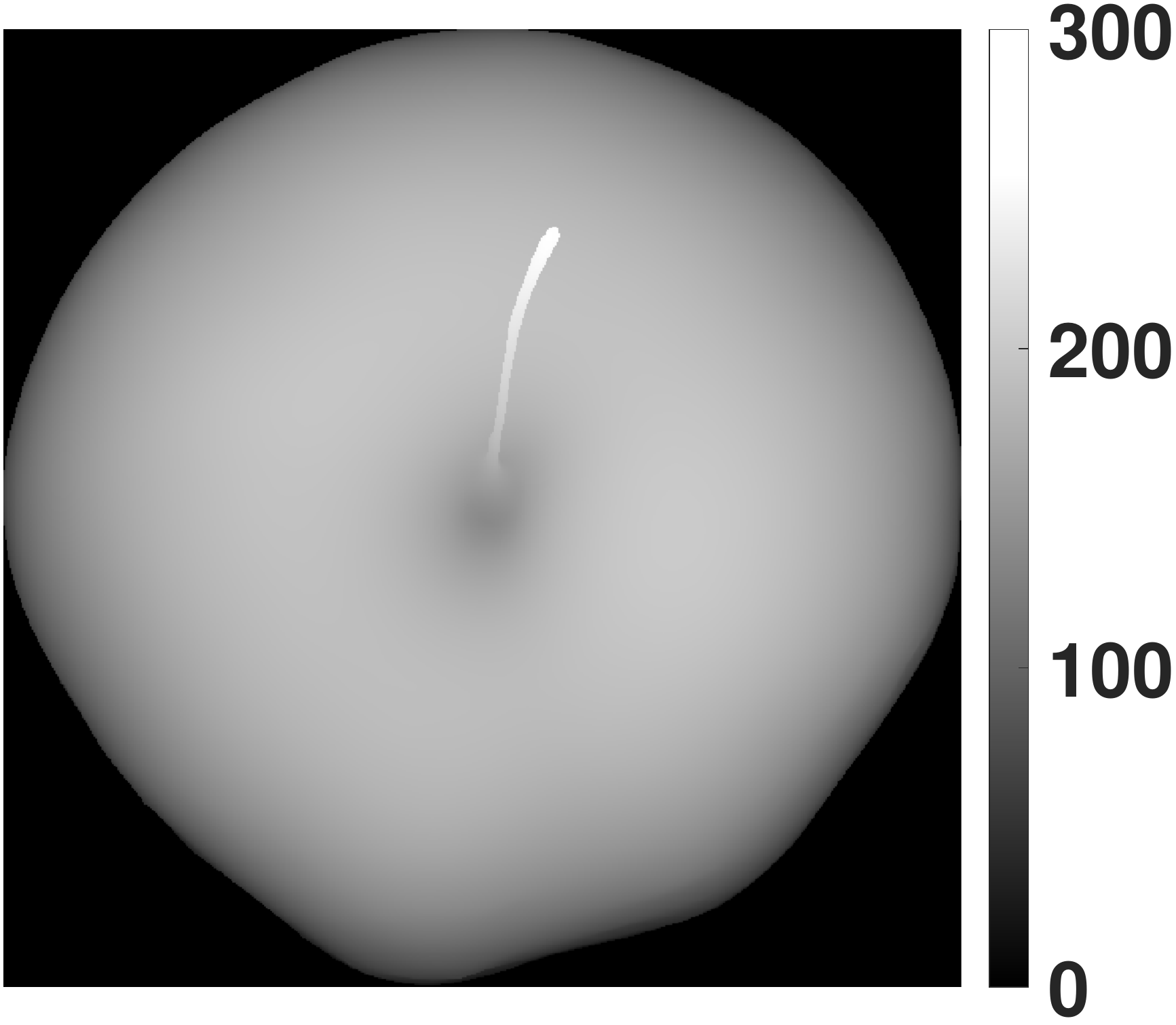}}
	\hspace{0.2em}
	\subfloat[\label{fig:experiments-apple-3D}]{\includegraphics[width=0.20\columnwidth]{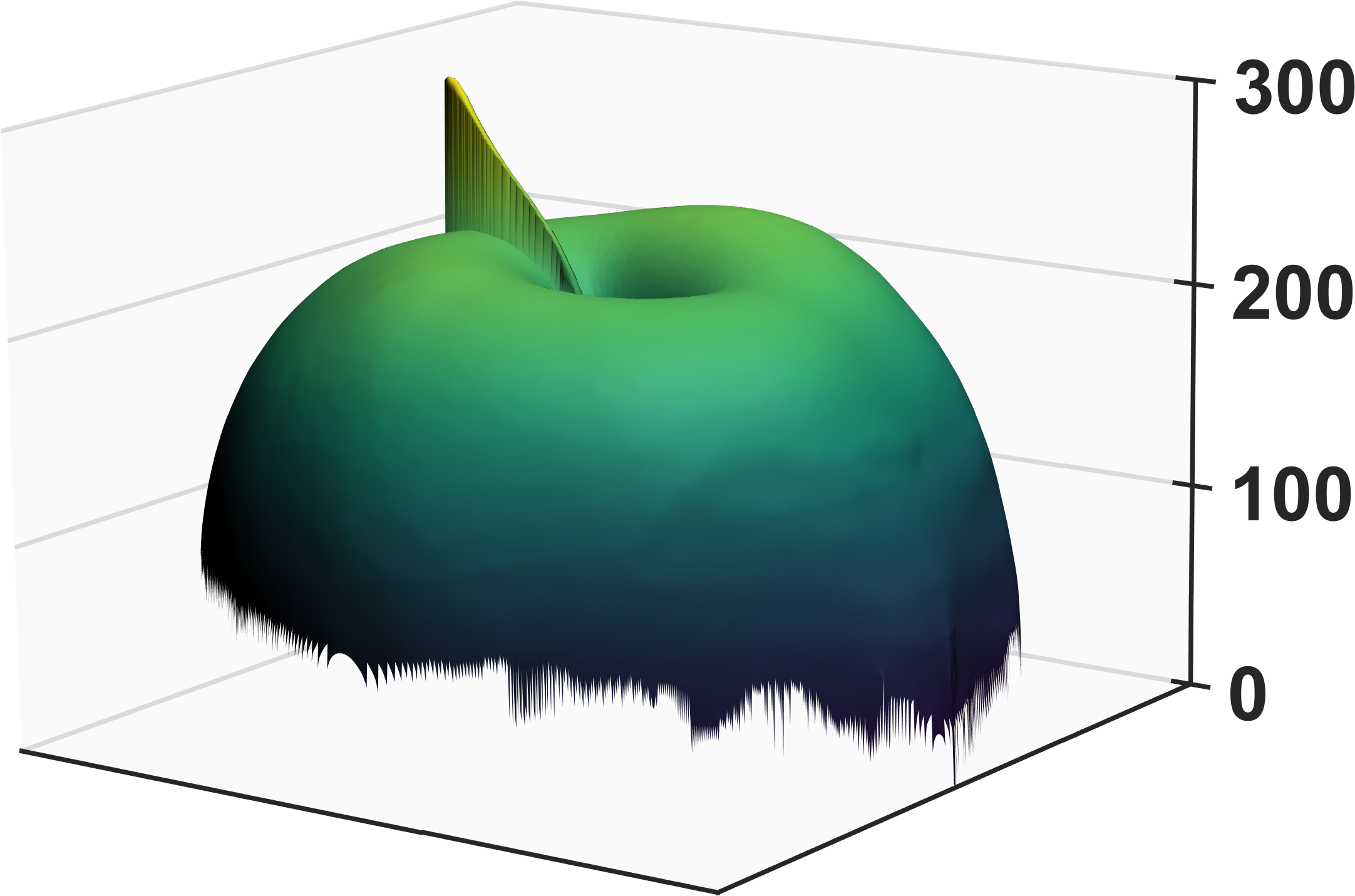}}
	\hspace{0.2em}
	\subfloat[\label{fig:experiments-apple-sphere}]{\includegraphics[width=0.20\columnwidth]{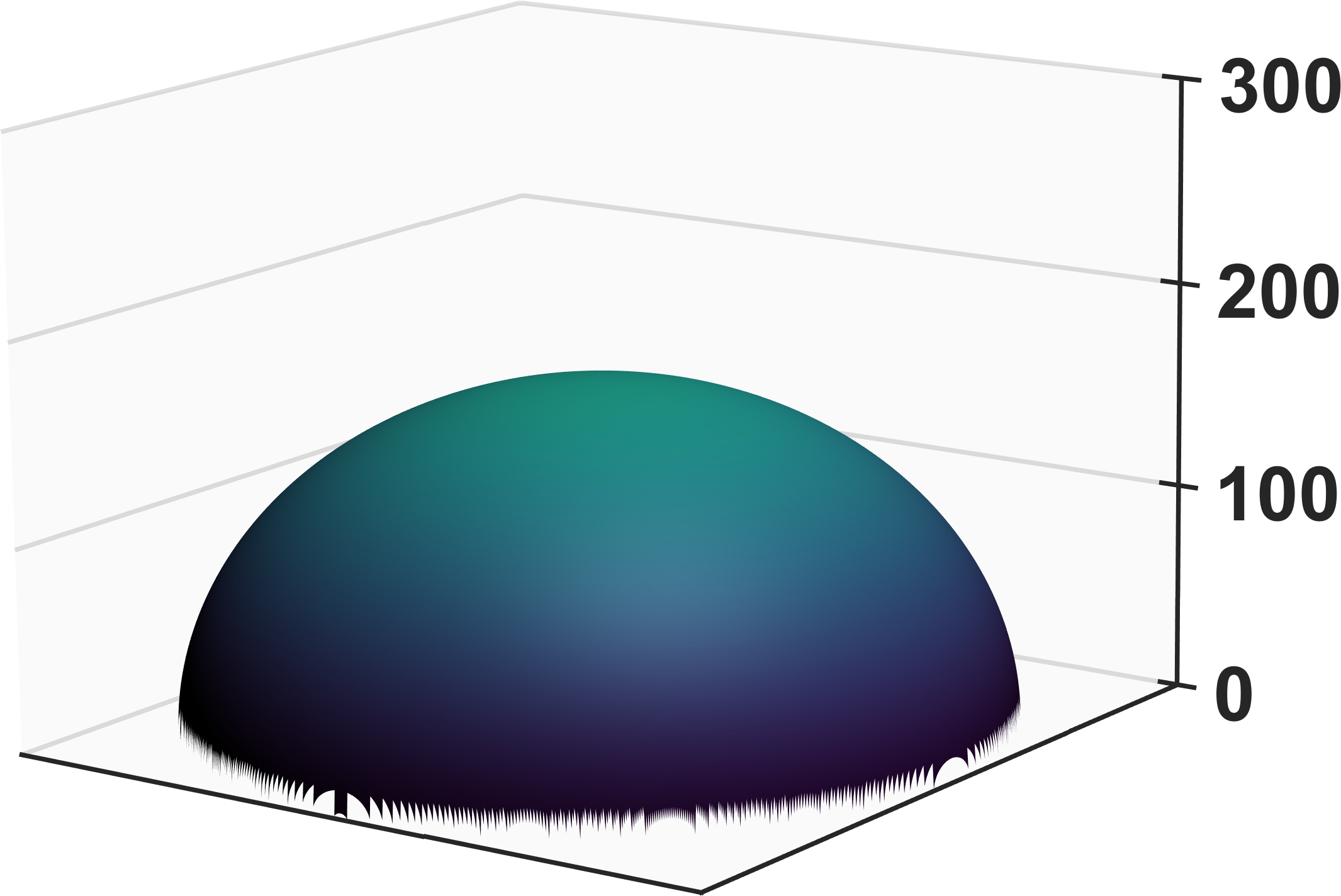}}
	\hspace{0.2em}
	\subfloat[\label{fig:experiments-apple-reduced}]{\includegraphics[width=0.20\columnwidth]{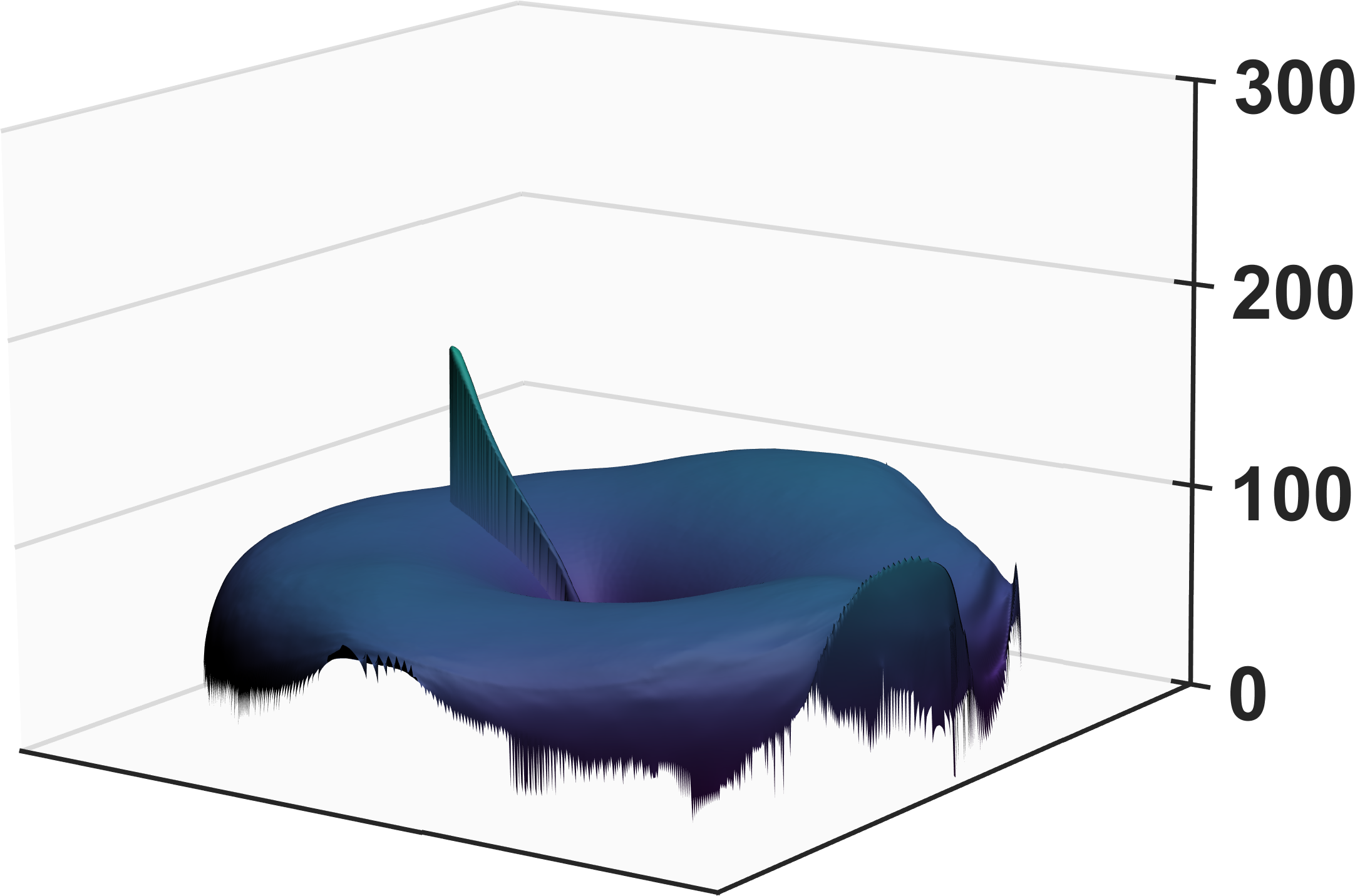}}
	\hspace{0.2em}
	\subfloat[\label{fig:experiments-apple-encoded}]{\includegraphics[width=0.13\columnwidth]{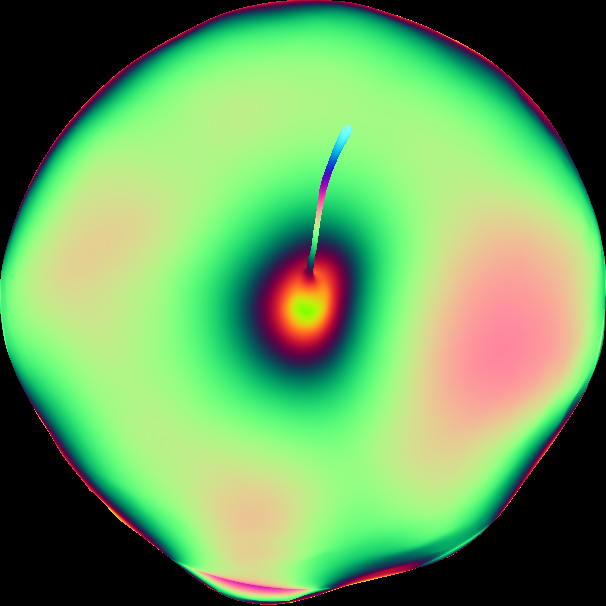}}
	
	\caption{Reducing depth range and approximation overhead using a primitive geometry. (a) the depth map, $Z$, to be encoded with a depth range of 297.7 mm; (b) 3D rendering of (a); (b) 3D rendering of $\widetilde{Z'}$, a primitive sphere that has been aligned and scaled to approximate $Z$; (c) 3D rendering of $Z_r$, the reduced range geometry, with a depth range of 165.8 mm; (e) the MWD-encoded image that stores $Z_r$.}
	
	\label{fig:experiments-apple}
\end{figure}

To reduce the amount of overhead needed to represent and regenerate $\widetilde{Z}'$, approximations may be made from primitive geometries.  
The final experiment in this paper highlights this flexibility of the proposed method.
Figure~\ref{fig:experiments-apple} is the 606 $\times$ 606 depth map, $Z$, of an apple model~\cite{Sketchfab:AppleModel} to be encoded.
Figures~\ref{fig:experiments-apple-depth} and \ref{fig:experiments-apple-3D} show the 2D and 3D renderings of depth data to be stored that has a depth range of 297.7 mm.
Figure~\ref{fig:experiments-apple-sphere} is a 3D rendering of $\widetilde{Z}'$, which was generated by fitting and scaling an ideal sphere to the apple's original 3D geometry.
This primitive sphere can be regenerated using only four parameters: a radius and the $x, y, z$ center coordinate of the sphere.
A 2D depth map is then generated from the 3D coordinates of the aligned sphere.
Figure~\ref{fig:experiments-apple-reduced} is a 3D rendering of $Z_r$, the range-reduced geometry produced by subtracting $\widetilde{Z}'$ from $Z$.
The depth range of $Z_r$ is 165.8 mm, 44.3\% less than the original depth range of 297.7 mm.
Figure~\ref{fig:experiments-apple-encoded} is the 2D encoded output image produced by the MWD encoding method with two encoding periods ($n$ = 2). 
The file size of the encoded image is 190.7 KB when stored in the PNG format, and $Z$ can be recovered with an RMS error of 0.0302 mm (99.99\% accuracy).
Since only four parameters are required to generate the approximated geometry ($\widetilde{Z}'$), the overhead storage cost is nearly negligible; even if floating-point precision is required to regenerate the approximation, only 16 bytes are needed to represent the sphere.
This experiment illustrates that if the original geometry can be well represented by a primitive, then overhead storage costs can be significantly reduced while still reducing the geometry's depth range prior to encoding.

\section{Discussion}
\label{sec:discussion}
The proposed method of depth range reduction is able to reduce the number of encoding periods required to compress 3D range geometry using state-of-the-art depth compression algorithms. 
This reduction in encoding frequency translates directly to reduced file sizes when storing and transmitting the compressed data.
There are, however, several important variables that must be addressed when applying this method to a set of data.
The following is a discussion of the main parameters that must be understood and appropriately tuned in order for the proposed method to achieve the desired effect. 

\begin{itemize}
	\item \textbf{Influence of approximation and alignment.} 
	Ideally, an approximation of the original geometry can be generated and aligned such that it can be subtracted from the original geometry to reduce the range of depth values present in a scene. 
	In reality, the approximation may be difficult to generate when the provided range geometry has surface holes (i.e., missing data), rapid variation near edges, or local regions with sharp depth changes (e.g., the apple stem in Fig.~\ref{fig:experiments-apple}).
	If the initial geometry has these features, it may also be difficult to precisely align the approximate geometry to the original.
	Overall, challenges in approximation and alignment can result in an insignificant depth range reduction or a reduction that leaves $Z_r$ with new high frequency features or rapidly varying structures.
	When $Z_r$ has these artifacts, the potential for file size savings via image-based compression is limited.
	
	\item \textbf{Limitations with lossy image-based compression.}
	In general, the proposed method can be described as an approach that removes low frequency components of a 3D range geometry scene, leaving only the high frequency features to be encoded.
	It follows that the encoding of $Z_r$ will have a higher spatial frequency than an equivalent encoding of the original geometry.
	This relatively higher spatial frequency results in larger file sizes and may cause compression artifacts as frequencies increase when using lossy compression to store the encoded image.
	Practically, this means that if a reduced range geometry has details of too high a frequency (i.e., too much of the low frequency information was removed), or the number of encoding periods is too large, then reconstructions from the encoded $Z_r$ may have an increased level of error.
	Despite these limitations, this paper has shown that it is possible to achieve file size savings when using lossy compression to store the encoded range-reduced 3D geometry.

	\item \textbf{Entropy reduction.}
	The experiments shown in this paper have assumed that the output of the proposed depth range reduction method is to be compressed using an image-based depth encoding method, such as MWD~\cite{Bell:MWD:2015}.
	If this encoding and compression step is not desired, the proposed method may still provide immediate file size savings by means of entropy reduction.
	More specifically, as the range of depth values decreases, so too does the number of bits required to represent the remaining values at some target precision.
	\begin{figure}[t!]
		\centering
		
		\includegraphics[width=0.99\columnwidth]{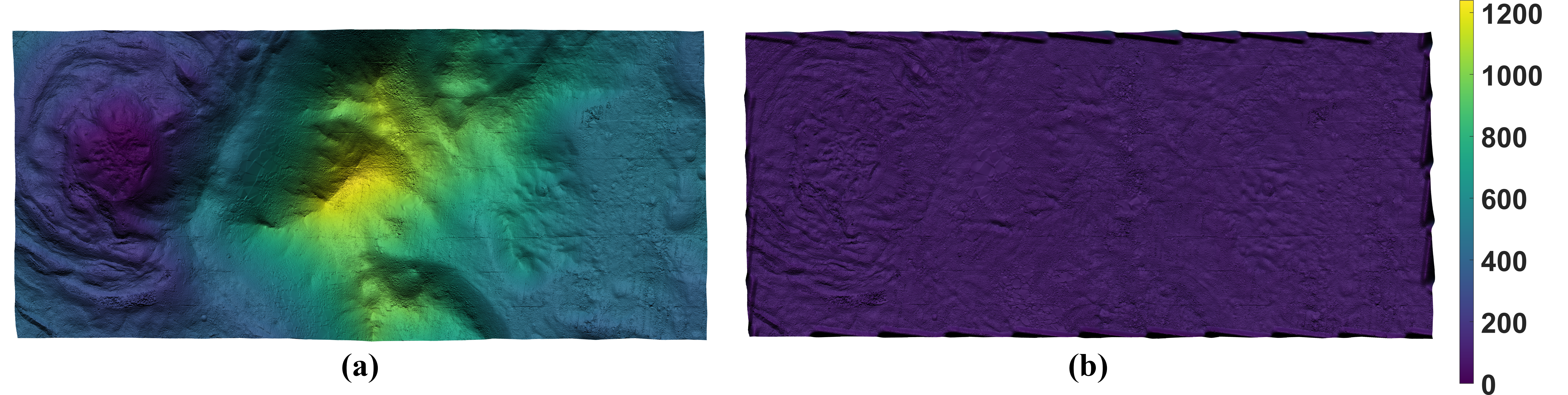}
		
		\caption{Applying the proposed depth range reduction method to a digital terrain model (DTM) captured by HiRISE (NASA/JPL/University of Arizona)~\cite{McEwen:HiRISE:2007}. (a) top-down 3D rendering of the DTM "Central Peak of an Impact Crater"~\cite{HiRISE:CentralPeakImpactCrater}; the data has a resolution of 5,556 $\times$ 2,996 with a depth range of approximately 1,239 meters. (b) top-down 3D rendering of the geometry in (a) after applying the proposed method; the range-reduced geometry has a new depth range of 553.8 meters.}
		
		\label{fig:discussion-hirise}
		
	\end{figure}
	Figure~\ref{fig:discussion-hirise} further illustrates this entropy reduction.
	Figure~\ref{fig:discussion-hirise}a is a 3D rendering of $Z$, the original geometry.
	In this case $Z$ is a $2,996 \times 5,556$ scan of the surface of Mars~\cite{HiRISE:CentralPeakImpactCrater} as captured by the Mars Reconnaissance Orbiter's High Resolution Science Experiment (HiRISE)~\cite{McEwen:HiRISE:2007}.
	An approximated geometry, $\widetilde{Z}'$, was generated by creating and resizing a $47 \times 87$ thumbnail image using the same 'blocking' approach employed in Sec.~\ref{sec:principle} and Sec.~\ref{sec:experiments}.
	Figure~\ref{fig:discussion-hirise}b is $Z_r$, the range-reduced geometry produced by subtracting $\widetilde{Z}'$ from $Z$.
	In this example the depth range is reduced from 1239.2 meters to 553.8 meters (55.3\% reduction in depth range).
	If the precision required by the target application is 0.1 meters, 14 bits per pixel are needed to represent values in the original geometry's depth range.
	In the reduced depth range, however, only 13 bits per pixel are required.
	In the raw format, this means that the original range geometry requires 28,447.4 KB to store and the reduced range geometry requires 26,419.4 KB to store (this value includes the overhead thumbnail image), resulting in a raw file size reduction of 2,028 KB (7.13\%).
	This example illustrates the ability of the proposed method to achieve entropy reduction in the raw format, regardless of the method used to compress the reduced geometry.
	
\end{itemize}

\section{Conclusion}
\label{sec:conclusion}
This paper has presented a novel method for the reduction of a 3D range geometry's depth range such that it can be encoded using a fewer number of encoding periods, resulting in a smaller output file size.
Specifically, the proposed method generates a low-overhead approximation of the original geometry and then subtracts it from the 3D range data to be compressed, effectively reducing the number of low frequency features within the scene.
Several methods of approximation were experimentally demonstrated, and potential challenges in the generation and alignment of the approximation were discussed.
It was also shown that the method allows for smaller file sizes to be achieved while maintaining some user-defined target reconstruction accuracy, enabling application-specific flexibility.
For example, the proposed method allowed for a 36.9\% reduction in file size to be achieved while targeting an RMS reconstruction error rate of 0.8 mm with JPEG 100.
Since the proposed method occurs prior to encoding, it is readily compatible with a variety of existing image-based 3D range geometry compression methods.
Alternatively, it was shown that the method could be used to simply reduce the number of bits per pixel required to store the geometry's raw data at a specified precision. 

\section*{Funding}
University of Iowa (ECE Department Faculty Startup Funds).

\section*{Acknowledgments}
The authors thank Broderick Schwartz for his feedback in the proofreading of this manuscript.


\bibliographystyle{plain}
\bibliography{2020_Depth-Range-Reduction}

\end{document}